\documentclass[
prd
,showpacs ,amssymb,nobibnotes,aps,eqsecnum]{revtex4}
\usepackage{graphicx}
\input{epsf}

\usepackage{amsmath,amssymb}
\usepackage{bm}

\newcommand{\dalm}{\kern1pt\vbox{\hrule height 0.9pt\hbox{\vrule width 0.9pt
\hskip 2.5pt\vbox{\vskip 5.5pt}\hskip 3pt\vrule width 0.3pt}\hrule height 0.3pt}
\kern1pt}

\begin{document}

%\twocolumn[\hsize\textwidth\columnwidth\hsize\csname @twocolumnfals\endcsname

% For two column
%\wideabs{

\title{Stellar Oscillations in Tensor-Vector-Scalar Theory}

\author{Hajime Sotani} \email{sotani@astro.auth.gr}
%\author{Paul Lasky$^{1,2,3}$} %\email{plasky@astro.swin.edu.au}
%\author{Dimitrios Giannios$^4$} %\email{giannios@MPA-Garching.MPG.DE}
\affiliation{
%
%Department of Physics, Aristotle University of Thessaloniki, 
%Thessaloniki 54124, Greece
%
Theoretical Astrophysics, University of T\"{u}bingen,
Auf der Morgenstelle 10, T\"{u}bingen 72076, Germany
%$^2$Centre for Astrophysics and Supercomputing, Swinburne University, Hawthorn VIC 3122, Australia \\
%$^3$Centre for Stellar and Planetary Astrophysics, Monash University, Clayton, VIC 3800, Australia \\
%$^4$Max Planck Institute for Astrophysics, Box 1317, D-85741 Garching, Germany
}

\date{\today}

% Abstract
\begin{abstract}
An alternative theory of gravity has recently been proposed by  
Bekenstein, named Tensor-Vector-Scalar (TeVeS) theory, which can  
explain many galactic and cosmological observations without the need  
for dark matter.  Whilst this theory passes basic solar system tests, and  
has been scrutinized with considerable detail in other weak-field  
regimes, comparatively little has been done in the strong-field limit  
of the theory. In this article, with Cowling approximation, we examine
the oscillation spectra of neutron stars in TeVeS.
As a result, we find that the frequencies of
fundamental modes in TeVeS could become lager than those expected
in general relativity, while the dependence of frequency of higher overtone
on gravitational theory is stronger than that of lower modes.
These imprints of TeVeS make it possible to distinguish the gravitational theory
in strong-field regime via the observations of gravitational waves,
which can provide unique confirmation of the existence of scalar field.
\end{abstract}

%
%%%%%%%%%%%%%%%%%%%%%%%%%%%%%%%%%%%%%%%%%%%%%%%%%%%%%%%%%%%%%%%
%  04.50.Kd :  Modified Theories of Gravity
%  04.80.Cc :  Experimental tests of gravitational theories
%  04.40.Dg :  Relativistic stars: structure, stability, and oscillations (see also 97.60.-s Late stages of stellar evolution) 
%  97.60.Jd :  Neutron stars (see also 26.60.+c Nuclear matter aspects of neutron stars in nuclear physics) 
%%%%%%%%%%%%%%%%%%%%%%%%%%%%%%%%%%%%%%%%%%%%%%%%%%%%%%%%%%%%%%%
%
%]
% For two column
%}
\maketitle
%\baselineskip 24pt
%%%%%%%%%%%%%%%%%%%%%%%%%%%%%%%%%%%%%%%%%%%%%%%%%%%%%%%%%%%%%%%%%%%%%%
%%%%%%%%%%%%%%%%%%%%%%%%%%%%%%%%%%%%%%%%%%%%%%%%%%%%%%%%%%%%%%%%%%%%%%
\section{Introduction}
\label{sec:I}
%%%%%%%%%%%%%%%%%%%%%%%%%%%%%%%%%%%%%%%%%%%%%%%%%%%%%%%%%%%%%%%%%%%%%%
%%%%%%%%%%%%%%%%%%%%%%%%%%%%%%%%%%%%%%%%%%%%%%%%%%%%%%%%%%%%%%%%%%%%%%

Tests of gravitational theories in the strong-field regime are extremely
important because, unlike the weak-field, they are still largely unconstrained by
observations.  
%There are still very few observations in the strong-field that are useful for
%dictating which is the appropriate theory of gravity.
However, with developing technology,
it is becoming possible to observe compact objects with high accuracy.
For example, observations of emitted X-rays and $\gamma$-rays
from compact objects could be used to directly test the strong-field regime of
a gravity theory \cite{Psaltis2008}. Furthermore, future observational
developments, for example with gravitational waves, will allow us to obtain
different physical properties for compact objects,
which will further allow for testing in the strong-field regime.  
From a theoretical point of view, there are attempts to test theories of gravity in the
strong-field regime by using surface atomic line redshifts \cite{DeDeo2003} or
gravitational waves from the neutron stars \cite{Sotani2004}.  In these
investigations, the possibility of distinguishing Scalar-Tensor (ST) theory, proposed
by Damour \& Esposito-Far\`ese \cite{Damour1992}, from General Relativity (GR) 
was discussed (see Psaltis \cite{Psaltis2008} for a review).  Whilst the
existence of scalar fields has not been experimentally verified, several experiments in the
weak-field limit of GR set severe limits on the existence and strength of such fields \cite{Will2001}.

Recently, Tensor-Vector-Scalar (TeVeS) theory has attracted considerable attention as an alternative
gravitational theory. TeVeS was proposed by Bekenstein \cite{Bekenstein2004}
as a relativistic theory for Modified Newtonian Dynamics \cite{Milgrom1983}. As such, it
explains galaxy rotation curves and the Tully-Fisher law without the existence of
dark matter.  TeVeS has also successfully explained strong gravitational
lensing \cite{Chen2006} as well as key features of the cosmic
microwave background \cite{Skordis2006} and galaxy distributions through an
evolving Universe \cite{Dodelson2006} without cold dark matter.  
While in the strong-field regime of TeVeS, Giannios
found the Schwarzschild solution \cite{Giannios2005}, and Sagi and
Bekenstein generalized this to the Reissner-Nordstr\"om solution \cite{Sagi2008}.
Furthermore, Contaldi, Wiseman \& Withers have found vacuum solutions
for a constant scalar field \cite{Contaldi2008}.
More recently, the Tolman-Oppenheimer-Volkoff (TOV) equations in TeVeS
are derived by Lasky, Sotani \& Giannios \cite{Paul2008},
with which one can produce static, spherically symmetric neutron stars,
and they showed the possibility of distinguishing TeVeS from GR by way of redshift observations.
In this article, we examine whether observations of gravitational waves associated with the
neutron star oscillations can provide an alternative way of probing the gravitational theory
in the strong-field regime.

The attempt to estimate the stellar parameters, such as
mass, radius and equation of state (EOS), via their oscillation properties is not a new idea.
Helioseismology is established fields in astronomy and one could know the information
about the interior of our sun. In the late '90s, it was suggested the possibility
to reveal the compact star properties through the oscillation spectra \cite{Andersson1996},
which is called as gravitational wave asteroseismology. The stellar mass, radius and EOS
can be deduced by an analysis of the oscillation spectrum of fundamental, pressure
and spacetime modes, i.e., $f$, $p$ and $w$ modes (e.g., \cite{Sotani2001,Sotani2003}).
While the rotation period of a compact star can be revealed by the examination of
$r$ mode oscillations (e.g., \cite{SPK2007,Miltos2008}),
where such frequencies are proportional to the rotation rate.
Furthermore, the detailed analysis of the gravitational waves makes it possible to
determine the radius of accretion disk around supermassive black hole \cite{Sotani2006} or
to know the magnetic effect during the stellar collapse \cite{SYK2007}.

In general, the oscillations of a neutron star in TeVeS could produce not only gravitational but also
scalar and vector waves, which is similar to the case in the scalar-tensor theory \cite{Will1993}, and
the direct detection of scalar and/or vector waves would be a unique probe for the gravitational theory.
Still, we will show that it is not necessary to detect these waves, because the obvious
imprints due to the existence of scalar and vector fields will be apparent in the spectrum of gravitational
waves associated with the stellar oscillations. Although we adopt Cowling approximation in this
article, the more complicated analysis including the metric, vector and scalar perturbations will be
seen near future somewhere.

This article is organized as follows. In the next section, we describe the fundamental parts of
TeVeS and TOV equations in TeVeS, where we also show
the neutron star models. In Sec. \ref{sec:III} we derive the perturbation equations with
Cowling approximation. Then the oscillation spectra of neutron stars in TeVeS are shown
in Sec. \ref{sec:IV}, finally we discuss the results related to gravitational wave asteroseismology
in Sec. \ref{sec:V}. In this article, we adopt the unit of $c=G=1$, where $c$
and $G$ denote the speed of light and the gravitational constant, respectively, and
the metric signature is $(-,+,+,+)$.

%%%%%%%%%%%%%%%%%%%%%%%%%%%%%%%%%%%%%%%%%%%%%%%%%%%%%
%%%%%%%%%%%%%%%%%%%%%%%%%%%%%%%%%%%%%%%%%%%%%%%%%%%%%
\section{Stellar Models in TeVeS}
\label{sec:II}
%%%%%%%%%%%%%%%%%%%%%%%%%%%%%%%%%%%%%%%%%%%%%%%%%%%%%
%%%%%%%%%%%%%%%%%%%%%%%%%%%%%%%%%%%%%%%%%%%%%%%%%%%%%
%%%%%%%%%%%%%%%%%%%%%%%%%%%%%%%%%%%%%%%%%%%%%%%%%%%%%
\subsection{TeVeS}
\label{sec:II-1}
%%%%%%%%%%%%%%%%%%%%%%%%%%%%%%%%%%%%%%%%%%%%%%%%%%%%%

Since details of TeVeS can be found in \cite{Bekenstein2004},
we only mention here the fundamental parts of the theory that are necessary for the present calculations.
TeVeS is based on three dynamical gravitational fields; an Einstein metric $g_{\mu\nu}$, 
a timelike 4-vector field ${\cal U}^\mu$,
and a scalar field $\varphi$.  There is also a nondynamical scalar field, $\sigma$.
The vector field fulfills the normalization condition,
$g_{\mu\nu}{\cal U}^\mu{\cal U}^\nu=-1$, and the physical metric is given by
\begin{gather}
 \tilde{g}_{\mu\nu} = e^{-2\varphi}g_{\mu\nu} - 2{\cal U}_\mu{\cal U}_\nu\sinh(2\varphi), \\
 \tilde{g}^{\mu\nu} = e^{2\varphi}g^{\mu\nu} + 2{\cal U}^\mu{\cal U}^\nu\sinh(2\varphi).
\end{gather}
All quantities in the physical frame are denoted with a tilde, and any
quantity without a tilde is in the Einstein frame.  The total action of TeVeS,
$S$, contains contributions from the three dynamical fields and a matter
contribution (see \cite{Bekenstein2004} for details).  These include two
positive dimensionless parameters, $k$ and $K$, which are the coupling
parameters for the scalar and vector fields respectively.  There also exists a
dimensionless free function $F$, a constant length scale $\ell$, and a spacetime dependent Lagrange multiplier, $\lambda$.

By varying the total action, %$S=S_g+S_s+S_v+S_m$ with respect to $g^{\mu\nu}$,
$S$, with respect to $g^{\mu\nu}$,
one can obtain the field equations for the tensor field
\begin{equation}
 G_{\mu\nu} = 8\pi G \left[\tilde{T}_{\mu\nu}+\left(1-e^{-4\varphi}\right){\cal U}^\alpha
               \tilde{T}_{\alpha(\mu}{\cal U}_{\nu)}+\tau_{\mu\nu}\right]+\Theta_{\mu\nu},
               \label{Einstein}
\end{equation}
where $\tilde{T}_{\mu\nu}$ is the energy-momentum
tensor in the physical frame, $\tilde{T}_{\alpha(\mu}{\cal U}_{\nu)}\equiv\tilde{T}_{\alpha\mu}{\cal U}_{\nu}
+\tilde{T}_{\alpha\nu}{\cal U}_{\mu}$ and $G_{\mu\nu}$ is the Einstein tensor
in the Einstein frame.  Conservation of energy-momentum is therefore given in
the physical frame as $\tilde{\nabla}_\mu \tilde{T}^{\mu\nu}=0$.
The other sources in Eq.(\ref{Einstein}) are given by
\begin{align}
 \tau_{\mu\nu} =& \sigma^2 \bigg[\varphi_{,\mu}\varphi_{,\nu}-\frac{1}{2}g^{\alpha\beta}
     \varphi_{,\alpha}\varphi_{,\beta}g_{\mu\nu} - \frac{G\sigma^2}{4\ell^2}F(kG\sigma^2)
     g_{\mu\nu}
     - {\cal U}^\alpha \varphi_{,\alpha}\left({\cal U}_{(\mu}\varphi_{,\nu)}
     -\frac{1}{2}{\cal U}^\beta\varphi_{,\beta}g_{\mu\nu}\right)\bigg], \label{tau} \\
 \Theta_{\mu\nu} =& K\left(g^{\alpha\beta}{\cal U}_{[\alpha,\mu]}{\cal U}_{[\beta,\nu]}
     - \frac{1}{4}g^{\gamma\delta}g^{\alpha\beta}{\cal U}_{[\gamma,\alpha]}{\cal U}_{[\delta,\beta]}
     g_{\mu\nu}\right)
     - \lambda {\cal U}_{\mu}{\cal U}_{\nu},
\end{align}
where ${\cal U}_{[\alpha,\beta]}\equiv{\cal U}_{\alpha,\beta}-{\cal U}_{\beta,\alpha}$.
Similarly, by varying $S$ with respect to ${\cal U}_\mu$ and $\varphi$,
one obtains the field equations for the vector and scalar fields;
\begin{gather}
% K\left({{\cal U}^{[\mu;\alpha]}}_{;\alpha} + {\cal U}^\mu {\cal U}_\alpha
%      {{\cal U}^{[\alpha;\beta]}}_{;\beta}\right) + 8\pi G\sigma^2 \left[{\cal U}^\alpha\varphi_{,\alpha}
%      g^{\mu\beta}\varphi_{,\beta} + {\cal U}^\mu\left({\cal U}^\alpha\varphi_{,\alpha}\right)^2\right]
%      = 8\pi G \left(1-e^{-4\varphi}\right)\left(g^{\mu\alpha}{\cal U}^\beta \tilde{T}_{\alpha\beta}
%      + {\cal U}^\mu{\cal U}^\alpha {\cal U}^\beta \tilde{T}_{\alpha\beta}\right), \label{vector} \\
 K{{\cal U}^{[\alpha;\beta]}}_{;\beta} + \lambda {\cal U}^\alpha + 8\pi G\sigma^2{\cal U}^\beta
      \varphi_{,\beta}g^{\alpha\gamma}\varphi_{,\gamma}
      = 8\pi G \left(1-e^{-4\varphi}\right)g^{\alpha\mu}
      {\cal U}^\beta\tilde{T}_{\mu\beta}, \label{vector} \\
 \left[\mu(k\ell^2h^{\mu\nu}\varphi_{,\mu}\varphi_{,\nu})h^{\alpha\beta}\varphi_{,\alpha}\right]_{;\beta}
      = kG\left[g^{\alpha\beta}+ \left(1+e^{-4\varphi}\right){\cal U}^\alpha{\cal U}^\beta\right]
      \tilde{T}_{\alpha\beta}, \label{scalar}
\end{gather}
where $h^{\alpha\beta}=g^{\alpha\beta}-{\cal U}^\alpha{\cal U}^\beta$ and $\mu(x)$ is a function defined by
$2\mu F(\mu) + \mu^2dF(\mu)/d\mu = -2x$.
With this function $\mu$, the nondynamical scalar field $\sigma$ is determined by
\begin{equation}
 kG\sigma^2 = \mu(k\ell^2h^{\alpha\beta}\varphi_{,\alpha}\varphi_{,\beta}). \label{scalar1}
\end{equation}
Therefore, the field equations of TeVeS are Eqs. (\ref{Einstein}) and (\ref{vector}) -- (\ref{scalar1}). 
%Now with the normalization condition for ${\cal U}^\mu$,
%from the vector equation (\ref{vector}) one can calculate the Lagrange multiplier $\lambda$.
%\begin{equation}
% \lambda = K {\cal U}_\alpha{{\cal U}^{[\alpha;\beta]}}_{;\beta} + \frac{8\pi}{k}
%           {\cal U}^{\alpha}{\cal U}^{\beta}\varphi_{,\alpha}\varphi_{,\beta}
%           - 8\pi G\left(1-e^{-4\varphi}\right){\cal U}^\alpha{\cal U}^\beta\tilde{T}_{\alpha\beta}.
%\end{equation}
%Since in this article we consider the strong gravitational region, we can put $\mu=1$.
It has been shown in the strong-field limit that $\mu=1$ is an excellent approximation \cite{Giannios2005,Sagi2008}.
On cosmological scales this is not a good choice \cite{Bekenstein2004}, however in this article
we only consider regions not too far from neutron stars, and we therefore set $\mu=1$.
%Notice that this selection for $\mu$ is valid even in the Newtonian limit,
%while on the cosmological region this is not good choice \cite{Bekenstein2004}.
%Thus we deal with the region which is not too far from the neutron stars.
This implies from Eq. (\ref{scalar1}) that $\sigma^2 = 1/(kG)$.
Moreover, whilst the functional form of $F$ is not predicted by the theory,
one can show that when $\mu=1$, the contribution of $F$ to the field equations
vanishes \cite{Bekenstein2004, Giannios2005, Sagi2008}.  Therefore, our
results are independent of this function and we drop it from the remaining discussion.

%%%%%%%%%%%%%%%%%%%%%%%%%%%%%%%%%%%%%%%%%%%%%%%%%%%%%
\subsection{TOV in TeVeS}
\label{sec:II-2}
%%%%%%%%%%%%%%%%%%%%%%%%%%%%%%%%%%%%%%%%%%%%%%%%%%%%%

First, the Tolman-Oppenheimer-Volkoff (TOV) equations in TeVeS are derived by Lasky, Sotani \& Giannios
\cite{Paul2008}. Here we make an brief description of TOV equations.
A static, spherically symmetric metric can be expressed as
\begin{equation}
 ds^2 = g_{\alpha\beta} dx^\alpha dx^\beta
      = -e^{\nu(r)} dt^2 + e^{\zeta(r)} dr^2 + r^2 d\Omega^2,
\end{equation}
where $d\Omega^2 =d\theta^2 + \sin^2\theta d\phi^2$ and $e^{-\zeta} = 1-2m(r)/r$.
In general, the vector field for a static, spherically symmetric spacetime can be described as
${\cal U}^\mu=\left({\cal U}^t,{\cal U}^r,0,0\right)$, where ${\cal U}^t$ and ${\cal U}^r$
are functions of $r$.  Giannios \cite{Giannios2005} showed that in vacuum, the
parameterized post-Newtonian (PPN) coefficients for a 
spherically symmetric, static spacetime
with a non-zero ${\cal U}^{r}$ can violate observational restrictions.  
In this article, we therefore only consider the case where ${\cal U}^r=0$.  In
this case, the vector field can be fully determined from the normalization
condition, and is found to be ${\cal U}^\mu = \left(e^{-\nu/2},0,0,0\right)$.  
Moreover, one can show that the vector field equation (\ref{vector}) is now trivially satisfied.
With this vector field, the physical metric is
\begin{equation}
 d\tilde{s}^2 = \tilde{g}_{\alpha\beta}dx^\alpha dx^\beta
              = -e^{\nu+2\varphi}dt^2 + e^{\zeta-2\varphi}dr^2
              + e^{-2\varphi}r^2d\Omega^2, \label{phys-metric}
\end{equation}
and the fluid four-velocity is $\tilde{u}_\mu = e^\varphi{\cal U}_\mu$.
We further assume the stellar matter content to be a perfect fluid, i.e.,
$\tilde{T}_{\mu\nu} = (\tilde{\rho} + \tilde{P})\tilde{u}_\mu\tilde{u}_\nu
                    + \tilde{P}\tilde{g}_{\mu\nu}$,
from which one can show that the full system of equations with $k\ne0$ and $K\ne 0$ reduces to
\begin{gather}
 \left(1-\frac{K}{2}\right)m'
            = \frac{Km}{2r} + 4\pi G r^2 e^{-2\varphi}\left(\tilde{\rho} + 2K\tilde{P}\right)
            + \left[\frac{2\pi r^2}{k}\psi^2 - \frac{Kr\nu'}{4}\left(1+\frac{r\nu'}{4}\right)\right]
              e^{-\zeta}, \\
 \frac{Kr}{4}\nu' = -1 + \left[1+K\left(\frac{4\pi Gr^3\tilde{P}e^{-2\varphi}+m}{r-2m}
            + \frac{2\pi r^2}{k}\psi^2\right)\right]^{1/2}, \\
 \tilde{P}' = -\frac{\tilde{P} + \tilde{\rho}}{2}(2\psi + \nu'), \label{eq:tov} \\
 \varphi'   = \psi, \\
 \psi'      = \left[\frac{m' r - m}{r(r-2m)}-\frac{\nu'}{2} - \frac{2}{r}\right]\psi
            + kGe^{-2\varphi+\zeta}\left(\tilde{\rho} + 3\tilde{P}\right),
\end{gather}
where a prime denotes a derivative with respect to $r$.
(See \cite{Paul2008} for the derivation of these equations and
for a discussion with $k=0$ and/or with $K=0$.)
This system of equations is closed with the addition of an equation of state (EOS).
The stellar radius in physical frame, $R$, is determined by $R\equiv e^{-\varphi(r_s)}r_s$,
where $r_s$ is the position of the stellar surface defined as the point where $\tilde{P}=0$.
Note that on exterior region the scalar field still exists although there is no fluid.

We integrate the above equations from the center, $r=0$, to the stellar surface, $r=r_s$. 
Moreover, the interior boundary conditions are given by $\tilde{P}(0)=\tilde{P}_0$,
$\tilde{\rho}(0)=\tilde{\rho}_0$, $\nu(0)=\nu_0$, $\varphi(0)=\varphi_0$, $\psi(0)=0$,
and $m(0)=0$, which are determined by Taylor series expansions of the above equations
near $r=0$ (see \cite{Paul2008} for details).
The exact values for $\nu_0$ and $\varphi_0$ are determined by matching the functions
$\nu(r)$ and $\varphi(r)$ to their asymptotic behavior,
which is found by performing a coordinate transformation on (\ref{phys-metric})
to bring it into an asymptotically flat form.  We define new coordinates
$\hat{t}\equiv te^{\varphi_{c}}$ and  $\hat{r}\equiv re^{-\varphi_{c}}$,
where $\varphi_{c}$ denotes the cosmological value of the scalar field.
Then performing an asymptotic expansion of all the equations and dropping the hats
on the new coordinates for simplicity in the expressions implies
\begin{align}
	\tilde{g}_{tt} &= -1+\frac{2M_{\rm ADM}}{r}+\mathcal{O}\left(\frac{1}{r^{2}}\right),\\
	\tilde{g}_{rr} &= 1+\frac{2M_{\rm ADM}}{r}+\mathcal{O}\left(\frac{1}{r^{2}}\right),\\
	\varphi &= \varphi_{c} - \frac{kGM_{\varphi}}{4\pi e^{\varphi_{c}}r}+\mathcal{O}\left(\frac{1}{r^{2}}\right).
\end{align}
Here, $M_{\rm ADM}$ is the total Arnowitt-Deser-Misner (ADM) mass given by
\begin{align}
	M_{\rm ADM}=\left(m_{\infty}+\frac{kGM_{\varphi}}{4\pi}\right)e^{-\varphi_{c}},
\end{align}
where $m_{\infty}$ is the mass function evaluated at radial infinity.  Also, $M_\varphi$ is the scalar mass \cite{Bekenstein2004}, which is constant outside the star and is defined everywhere as
\begin{equation}
 M_\varphi = 4\pi\int_0^{r}r^2\left(\tilde{\rho}+3\tilde{P}\right)e^{(\nu+\zeta)/2-2\varphi}dr.
\end{equation}
%
%
%where $r_s$ is the position of stellar surface.

We adopt the same EOS as in \cite{Sotani2004}, which are polytropic ones
derived by fitting functions to tabulated data of realistic EOS known as 
EOS A and EOS II. The maximum masses of neutron stars in GR
are $M=1.65 M_{\odot}$ with $R = 8.9$ km for EOS A and $M=1.95 M_{\odot}$ with
$R=10.9$ km for EOS II.  That is, EOS A is considered soft, while EOS II is an intermediate EOS.

When it comes to the study of the structure of neutron stars, TeVeS introduces three new 
parameters, $k$, $K$, and $\varphi_c$, with respect to GR.
Since the value of $k$ is tightly constrained by both cosmological models and also
planetary motions in the outer solar system \cite{Bekenstein2004},  we
accordingly set $k=0.03$ for the remainder of the article. 
%Moreover, we can show that the dependence on $\varphi_{c}$ is minimal for neutron star models. 
With respect to the value of $\varphi_c$,  Lasky, Sotani \& Giannios
showed that $\varphi_{c}$ can have a minimum value of around $0.001$,
based on causality issues inside the neutron star  \cite{Paul2008}.
Therefore, for this article we use $\varphi_{c}=0.003$.  Details of 
neutron star models where these parameters are allowed to vary are given in 
Lasky, Sotani \& Giannios \cite{Paul2008}, in which
they showed that the dependences on $k$ and on $\varphi_{c}$ are minimal for neutron star models.
While, restrictions on $K$ are less
severe, and have not been discussed in great detail in the literature.  In
this article we consider the range $0<K<2$, because for $K>2$ one can show that 
the pressure diverges from the stellar center outward, and therefore stellar models
are not possible \cite{Paul2008} (Sagi \& Bekenstein \cite{Sagi2008} also
showed that physical black hole solutions are only valid for $K<2$).

%%%%%%%%%%%%%%%%%%%%%%%%%%%%%%%%%%%%%%%%%%%%%%%%%%%%%
\subsection{Neutron Star Models in TeVeS}
\label{sec:II-3}
%%%%%%%%%%%%%%%%%%%%%%%%%%%%%%%%%%%%%%%%%%%%%%%%%%%%%

\begin{figure}[htbp]
\begin{center}
\begin{tabular}{cc}
\includegraphics[scale=0.45]{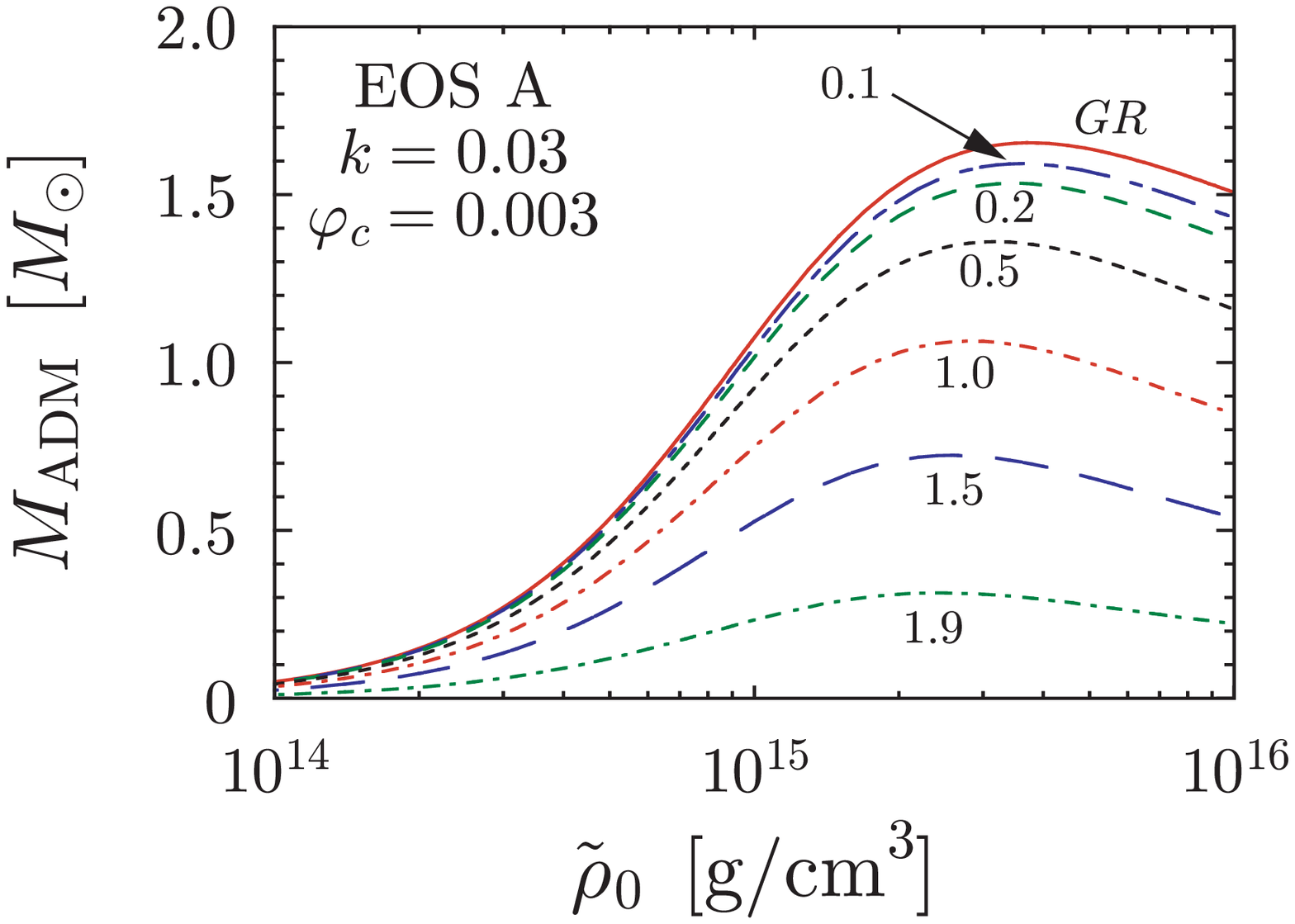} &
\includegraphics[scale=0.45]{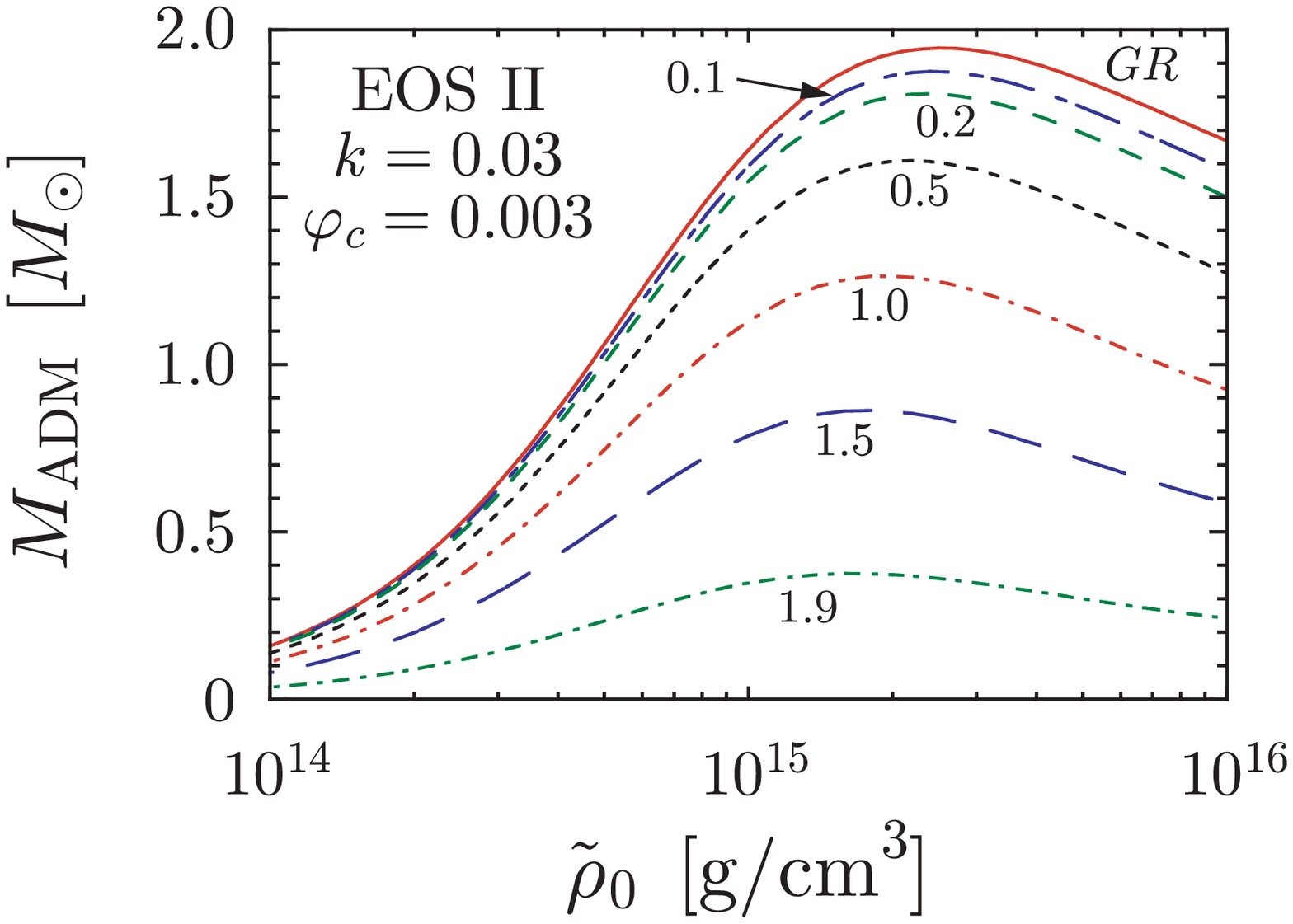} \\
\end{tabular}
\end{center}
\caption{%%
Relation between the mass and central density of neutron stars in GR and
in TeVeS with $k=0.03$, $\varphi_c=0.003$, where the left and right panels
are corresponding to the stellar properties given
by EOS A and EOS II, respectively. In the figure the solid
line denotes the case of GR while the other lines are
corresponding to the stellar models with different values of
$K$ in TeVeS.
}%%
\label{fig:rhoc-M}
\end{figure}

Fig. \ref{fig:rhoc-M} shows the ADM mass as a function of the
central density of neutron stars. Different lines correspond to different values of $K$, whose
values are indicated. Additionally,
we plot the stellar model for GR with the solid line.
Note that for spherically symmetric neutron stars,
stellar models for the region $\partial M_{\rm ADM}/\partial \tilde{\rho}_0<0$
could be unstable. From this figure we can see that, 
although the central density giving the maximum mass
is almost independent from the value of $K$, the corresponding maximum
mass depends strongly on the parameter $K$, i.e., for larger values of
$K$ the maximum mass becomes smaller. For example, with EOS A for $K=0.5$ the
maximum mass is 18 \% smaller than that of GR.
%However, given that the observed masses of neutron stars are larger than $1.27
%M_{\odot}$ \cite{Bhat2008}, we can rule out models at least where $K\gsim 1$.
%It is difficult to rule out an exact parameter space due to uncertainties in the EOS.

%%%%%%%%%%%%%%%%%%%%%%%%%%%%%%%%%%%%%%%%%%%%%%%%%%%%%%%%%%%%%%%%%
%  FIGURE
%%%%%%%%%%%%%%%%%%%%%%%%%%%%%%%%%%%%%%%%%%%%%%%%%%%%%
% Figure 1
%
%%%%%%%%%%%%%%%%%%%%%%%%%%%%%%%%%%%%%%%%%%%%%%%%%%%%%
%%%%%%%%%%%%%%%%%%%%%%%%%%%%%%%%%%%%%%%%%

\begin{figure}[htbp]
\begin{center}
\begin{tabular}{cc}
\includegraphics[scale=0.45]{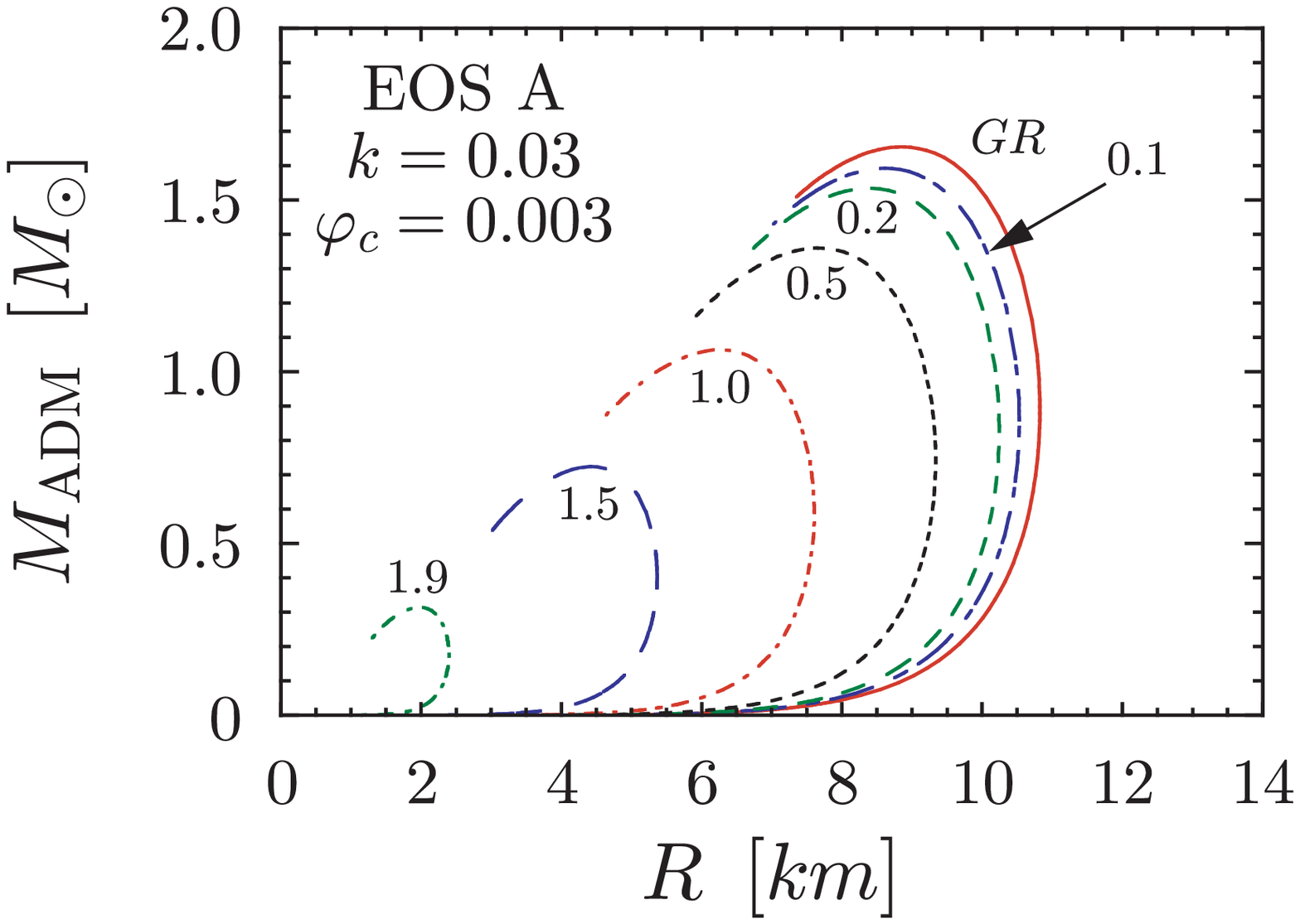} &
\includegraphics[scale=0.45]{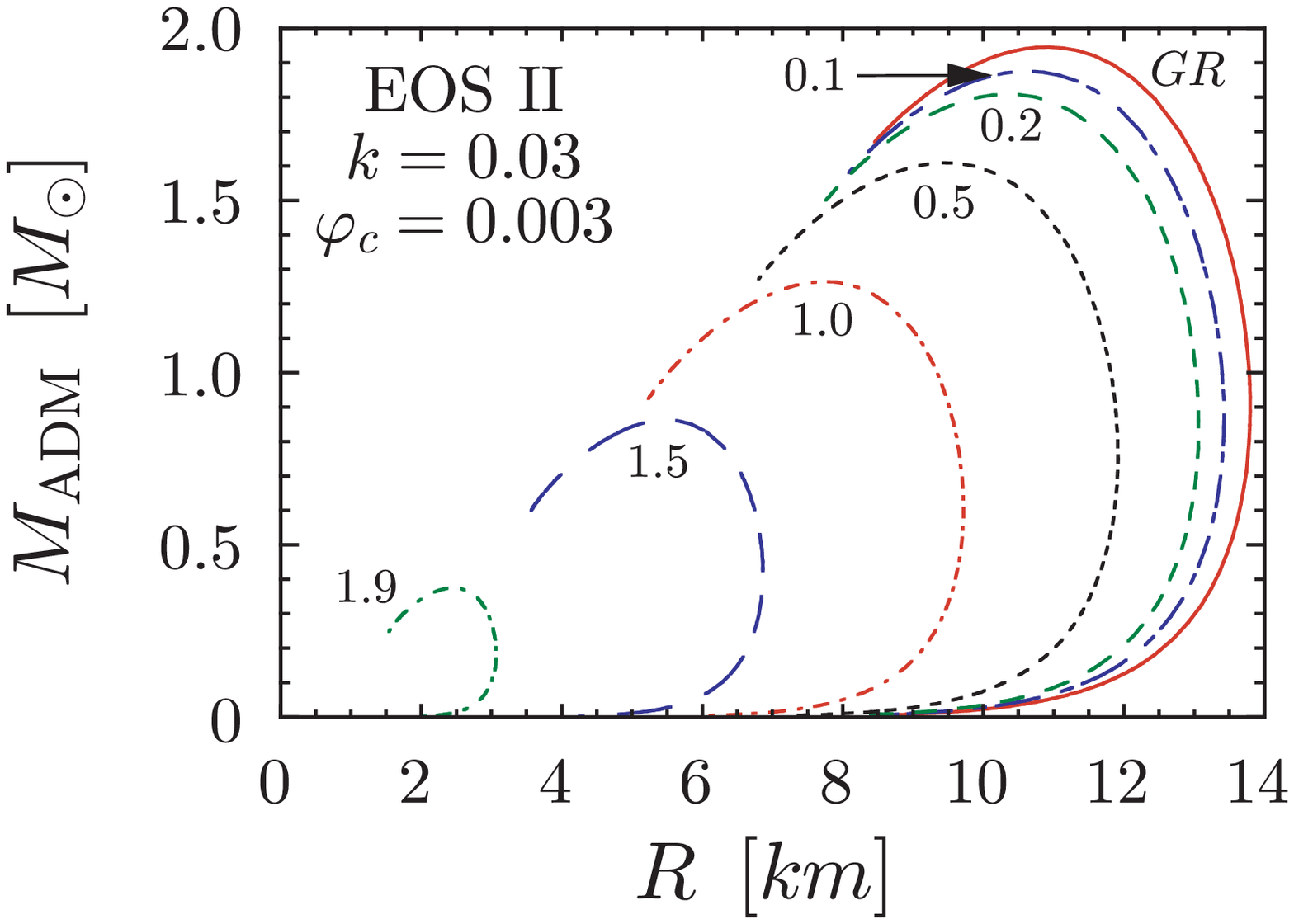} \\
\end{tabular}
\end{center}
\caption{%%
Mass-radius relation for neutron stars in GR and in TeVeS with $k=0.03$, $\varphi_c=0.003$,
where the left and right panels are corresponding to the stellar properties given
by EOS A and EOS II, respectively.
}%%
\label{fig:R-M}
\end{figure}

Fig. \ref{fig:R-M} shows the relation between the ADM mass and stellar radius
%, $R\equiv e^{-\varphi(r_s)} r_s$ (where $r_s$ is the position of the stellar surface),
with different values of $K$ and also for the GR case.  
In general, one requires a softer EOS near the stellar surface,
which implies the stellar radius becomes larger, but with high central density
for typical neutron star, the stellar models are almost independent from the consideration
of softer EOS near the stellar surface.  
This figure further implies that there exists a minimum radius,
which usually corresponds to the maximum mass \cite{Lattimer2001}.
Considering this minimum radius, it can be seen in Fig. \ref{fig:R-M}
that neutron stars in TeVeS are smaller than in the GR case.
For example, we can see that the minimum radius for a star with EOS A in TeVeS with
$K=0.5$ is $7.7$ km, whereas for GR it is $8.9$ km, which is a
$13.5\%$ difference.

%%%%%%%%%%%%%%%%%%%%%%%%%%%%%%%%%%%%%%%%%%%%%%%%%%%%%%%%%%%%%%%%%
%  FIGURE
%%%%%%%%%%%%%%%%%%%%%%%%%%%%%%%%%%%%%%%%%%%%%%%%%%%%%
% Figure 2
%
%%%%%%%%%%%%%%%%%%%%%%%%%%%%%%%%%%%%%%%%%%%%%%%%%%%%%
%%%%%%%%%%%%%%%%%%%%%%%%%%%%%%%%%%%%%%%%%

%%%%%%%%%%%%%%%%%%%%%%%%%%%%%%%%%%%%%%%%%%%%%%%%%%%%%%%%%%%%%%%%%
%%%%%%%%%%%%%%%%%%%%%%%%%%%%%%%%%%%%%%%%%%%%%%%%%%%%%%%%%%%%%%%%%
\section{Perturbation equations in the Cowling Approximation}
\label{sec:III}
%%%%%%%%%%%%%%%%%%%%%%%%%%%%%%%%%%%%%%%%%%%%%%%%%%%%%%%%%%%%%%%%%

In this section we derive the perturbation equations for nonradial oscillations
of spherically symmetric neutron stars in TeVeS. For simplicity, we adopt the Cowling approximation,
in which the fluid is perturbed on a fixed background. That is, the perturbations of the spacetime,
vector field and scalar field are frozen, i.e., $\delta \tilde{g}_{\mu\nu}=0$, $\delta {\cal U}^\mu=0$
and $\delta \varphi=0$.
It notes that with the Cowling approximation we can study only the oscillation modes related to the
fluid perturbations, such as $f$, $p$ and $g$ modes, while it is impossible to study the other emissions
of scalar waves, vector waves and gravitational waves connected to the oscillation of spacetime.
Additionally, we should notice that the Cowling approximation in GR is typically very good for axial type
of oscillations while for polar type of oscillations the error for typical relativistic stellar models 
could become less than 20 per cent for $f$ modes and around 10 per cent for $p$ modes \cite{Shijun1997}.

With Cowling approximation, the perturbed energy-momentum tensor, $\delta\tilde{T}^{\mu\nu}$,
is given as
\begin{equation}
 \delta \tilde{T}^{\mu\nu} = \left(\delta\tilde{\rho} + \delta\tilde{P}\right)\tilde{u}^\mu\tilde{u}^\nu
             + \left(\tilde{\rho} + \tilde{P}\right)
               \left(\delta\tilde{u}^\mu\tilde{u}^\nu + \tilde{u}^\mu\delta\tilde{u}^\nu\right)
             + \delta\tilde{P}\tilde{g}^{\mu\nu}.
\end{equation}
Introducing the Lagrangian displacement vector, the perturbed variables in $\delta\tilde{T}^{\mu\nu}$
such as $\delta\tilde{u}^\mu$, $\delta\tilde{\rho}$ and $\delta\tilde{P}$, can be described explicitly.
The Lagrangian displacement vector for the fluid perturbations are
\begin{align}
 \tilde{\xi}^i = \left(\tilde{\xi}^r,\tilde{\xi}^\theta,\tilde{\xi}^\phi\right)
               = \left(W,-V\partial_\theta,-V\sin^{-2}\theta\partial_\phi\right)\frac{1}{r^2}Y_{\ell m},
\end{align}
where $W$ and $V$ are functions of $t$ and $r$.
Then the perturbations of 4-velocity, $\delta \tilde{u}^\mu$, can be written as
\begin{align}
 \delta \tilde{u}^\mu = \left(0,\dot{W}, -\dot{V} \partial_\theta, -\dot{V} \sin^{-2}\theta \partial_\phi\right)
                                e^{-\varphi-\nu/2}\frac{1}{r^2}Y_{\ell m},
\end{align}
where dots on the variables denote the partial derivative with respect to $t$.
On the other hand, using the first law of thermodynamics, we can get the following relation
between the adiabatic changes of the density and the baryon number density;
\begin{equation}
 \Delta \tilde{\rho} = \frac{\tilde{\rho} + \tilde{P}}{\tilde{n}}\Delta \tilde{n},
     \label{eq:1st-law}
\end{equation}
where $\tilde{n}$ denotes the baryon number density.
So if we use the relationship between the Lagrangian perturbation $\Delta\tilde{\rho}$
and Eulerian perturbation $\delta\tilde{\rho}$ such as
\begin{equation}
 \Delta \tilde{\rho}
        %= \tilde{\rho}(t,r+\tilde{\xi}^r) - \tilde{\rho}(t,r)
    \simeq \delta \tilde{\rho} + \tilde{\xi}^r\partial_r\tilde{\rho},
\end{equation}
we can express the Eulerian density variation as
\begin{equation}
 \delta \tilde{\rho} = (\tilde{\rho} + \tilde{P})\frac{\Delta \tilde{n}}{\tilde{n}}
                     - \frac{\tilde{\rho}'W}{r^2}Y_{\ell m}.
\end{equation}
Additionally, with the definition of the adiabatic constant
\begin{equation}
 \gamma \equiv \left(\frac{\partial\ln\tilde{P}}{\partial\ln\tilde{n}}\right)_s
             = \frac{\tilde{n}\Delta\tilde{P}}{\tilde{P}\Delta\tilde{n}},
     \label{eq:gamma-def}
\end{equation}
we can derive the Eulerian variation of the pressure;
\begin{equation}
 \delta \tilde{P} = \gamma \tilde{P}\frac{\Delta \tilde{n}}{\tilde{n}} - \frac{\tilde{P}'W}{r^2}Y_{\ell m}.
\end{equation}
It notes that with Eqs. (\ref{eq:1st-law}) and (\ref{eq:gamma-def})
we can get the useful expression for $\gamma$ as
\begin{equation}
 \gamma = \frac{\tilde{\rho} + \tilde{P}}{\tilde{P}}\left(\frac{\partial \tilde{P}}{\partial\tilde{\rho}}\right)_s.
\end{equation}
Finally, the Lagrangian variation of the baryon number density, which comes on the expressions of
$\delta\tilde{\rho}$ and $\delta\tilde{P}$, is determined by the relation as
\begin{equation}
 \frac{\Delta\tilde{n}}{\tilde{n}}
      = -\tilde{\nabla}_k^{(3)}\tilde{\xi}^k - \frac{\delta\tilde{g}}{2\tilde{g}},
\end{equation}
where $\tilde{\nabla}_k^{(3)}$ and $\tilde{g}$ denote the covariant derivative in a 3-dimension
with metric $\tilde{g}_{\mu\nu}$ and the determinant of $\tilde{g}_{\mu\nu}$, respectively.
In this article, since we assume the Cowling approximation the second term is neglected. Then
the Lagrangian variation of the baryon number density can be written as
\begin{equation}
 \frac{\Delta \tilde{n}}{\tilde{n}} = -\left[W' + \frac{1}{2}\left(\zeta' - 6\varphi'\right)W
           + \ell(\ell + 1)V\right]\frac{1}{r^2}Y_{\ell m}.
\end{equation}

Finally we can get the equations describing the fluid perturbations by taking a variation of the energy-momentum
conservation law, $\tilde{\nabla}_\nu\tilde{T}^{\mu\nu}=0$. With Cowling approximation, this equation becomes
$\tilde{\nabla}_\nu\delta\tilde{T}^{\mu\nu}=0$. The explicit forms with $\mu=r,\theta$ are
\begin{align}
 \partial_r\left[\frac{\gamma\tilde{P}}{r^2}\left\{W'+\frac{1}{2}\left(\zeta'-6\varphi'\right)W
     + \ell(\ell+1)V\right\}+\frac{\tilde{P}'W}{r^2}\right] - \left(\tilde{\rho}+\tilde{P}\right)
       e^{-4\varphi-\nu+\zeta}\frac{\ddot{W}}{r^2} \nonumber \\
     - \frac{\tilde{\rho}' + \tilde{P}'}{\tilde{\rho} + \tilde{P}}
       \left[\frac{\gamma\tilde{P}}{r^2}\left\{W'+\frac{1}{2}\left(\zeta'-6\varphi'\right)W
     + \ell(\ell+1)V\right\}+\frac{\tilde{P}'W}{r^2}\right] = 0, \label{eq:perturbation1} \\
 \left(\tilde{\rho} + \tilde{P}\right)e^{-4\varphi-\nu}\ddot{V}
     + \frac{\gamma\tilde{P}}{r^2}\left\{W'+\frac{1}{2}\left(\zeta'-6\varphi'\right)W
     + \ell(\ell+1)V\right\}+\frac{\tilde{P}'W}{r^2} = 0,  \label{eq:perturbation2}
\end{align}
where we use the relation that $\delta\tilde{\rho}/\delta\tilde{P}=\tilde{\rho}'/\tilde{P}'$ and Eq. (\ref{eq:tov}).
By assuming a harmonic dependence on time, the perturbative variables will be written as $W(t,r)=W(r)e^{i\omega t}$
and $V(t,r)=V(r)e^{i\omega t}$. To make the above equations simpler, by calculating the combination of the form
$d(\ref{eq:perturbation2})/dr-(\ref{eq:perturbation1})$ and substituting Eq. (\ref{eq:perturbation2}) again,
we can get
\begin{equation}
 V' = \left(4\varphi' + \nu'\right)V - e^{\zeta}\frac{W}{r^2}. \label{eq:perturbation3}
\end{equation}
Thus, from Eqs. (\ref{eq:perturbation2}) and (\ref{eq:perturbation3}), we can obtain the following simple equation
system for the perturbations of fluid;
\begin{align}
 W' &= \frac{d\tilde{\rho}}{d\tilde{P}}\left[\omega^2r^2e^{-4\varphi-\nu}V
     + \frac{1}{2}\left(2\varphi' + \nu'\right)W\right] + \frac{1}{2}\left(6\varphi'-\zeta'\right)W
     - \ell(\ell+1)V, \label{eq:master1} \\
 V' &= \left(4\varphi' + \nu'\right)V - e^{\zeta}\frac{W}{r^2}. \label{eq:master2}
\end{align}
With the appropriate boundary conditions at the center and stellar surface, the above equation system constitutes
an eigenvalue problem for the parameter $\omega$. One can find the behavior of $W$ and $V$ near the stellar
center as $W(r)=Br^{\ell+1}+{\cal O}(r^{\ell+3})$ and $V(r)=-B r^\ell/\ell + {\cal O}(r^{\ell+2})$,
where $B$ is an arbitrary constant, while the boundary condition at the stellar surface is the vanishing the
Lagrangian perturbation of the pressure, i.e., $\Delta\tilde{P}=0$. Since the Lagrangian perturbation of
the pressure is described by $\Delta\tilde{P}=\gamma\tilde{P}\Delta\tilde{n}/\tilde{n}$, with a help of
Eq. (\ref{eq:master1}) we can get the boundary condition at the stellar surface as
\begin{equation}
 2\omega^2r^2e^{-4\varphi-\nu}V + \left(2\varphi' + \nu'\right)W = 0.
\end{equation}

%%%%%%%%%%%%%%%%%%%%%%%%%%%%%%%%%%%%%%%%%%%%%%%%%%%%%
%%%%%%%%%%%%%%%%%%%%%%%%%%%%%%%%%%%%%%%%%%%%%%%%%%%%%
\section{Oscillation Spectra}
\label{sec:IV}
%%%%%%%%%%%%%%%%%%%%%%%%%%%%%%%%%%%%%%%%%%%%%%%%%%%%%

\begin{figure}[htbp]
\begin{center}
\begin{tabular}{cc}
\includegraphics[scale=0.45]{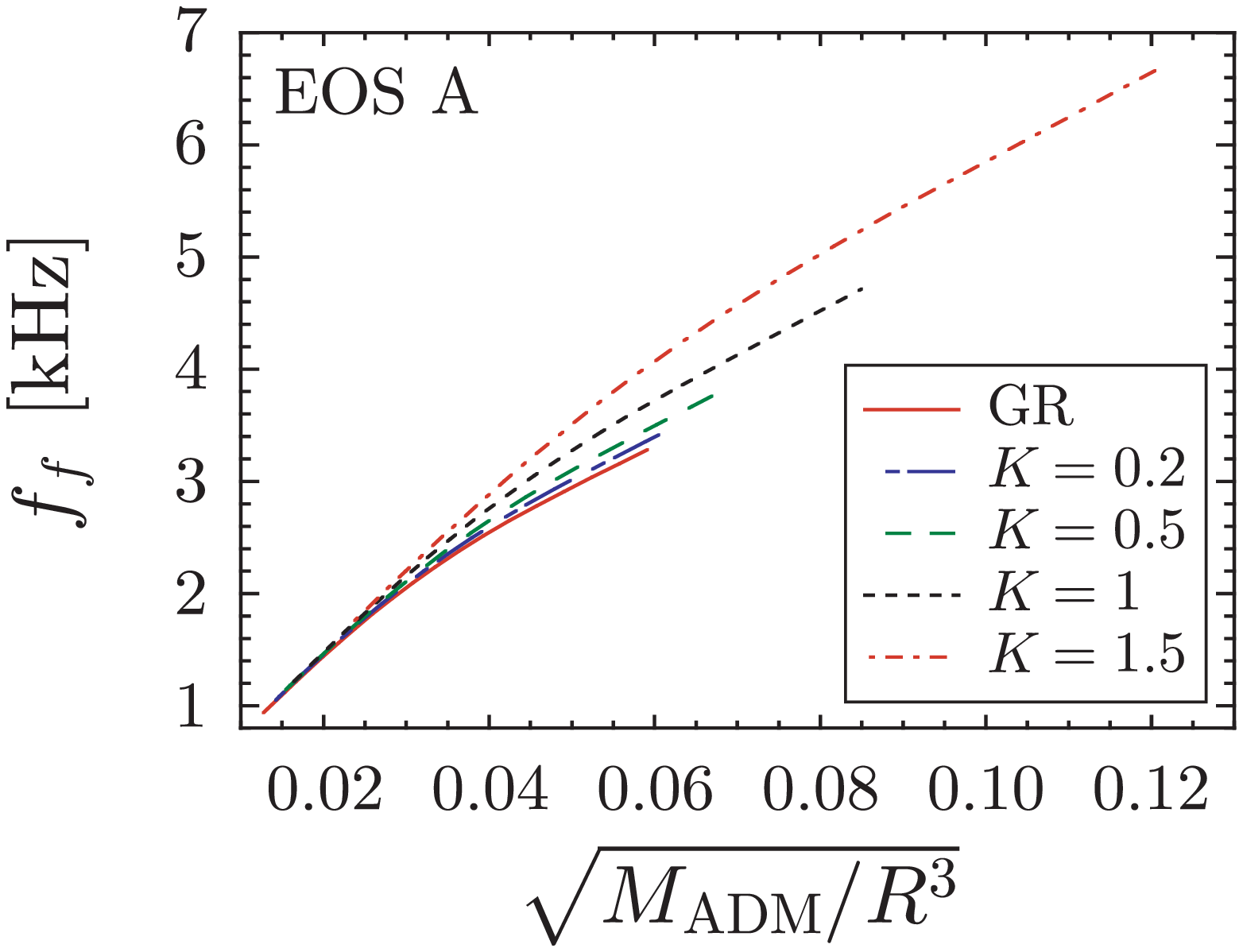} &
\includegraphics[scale=0.45]{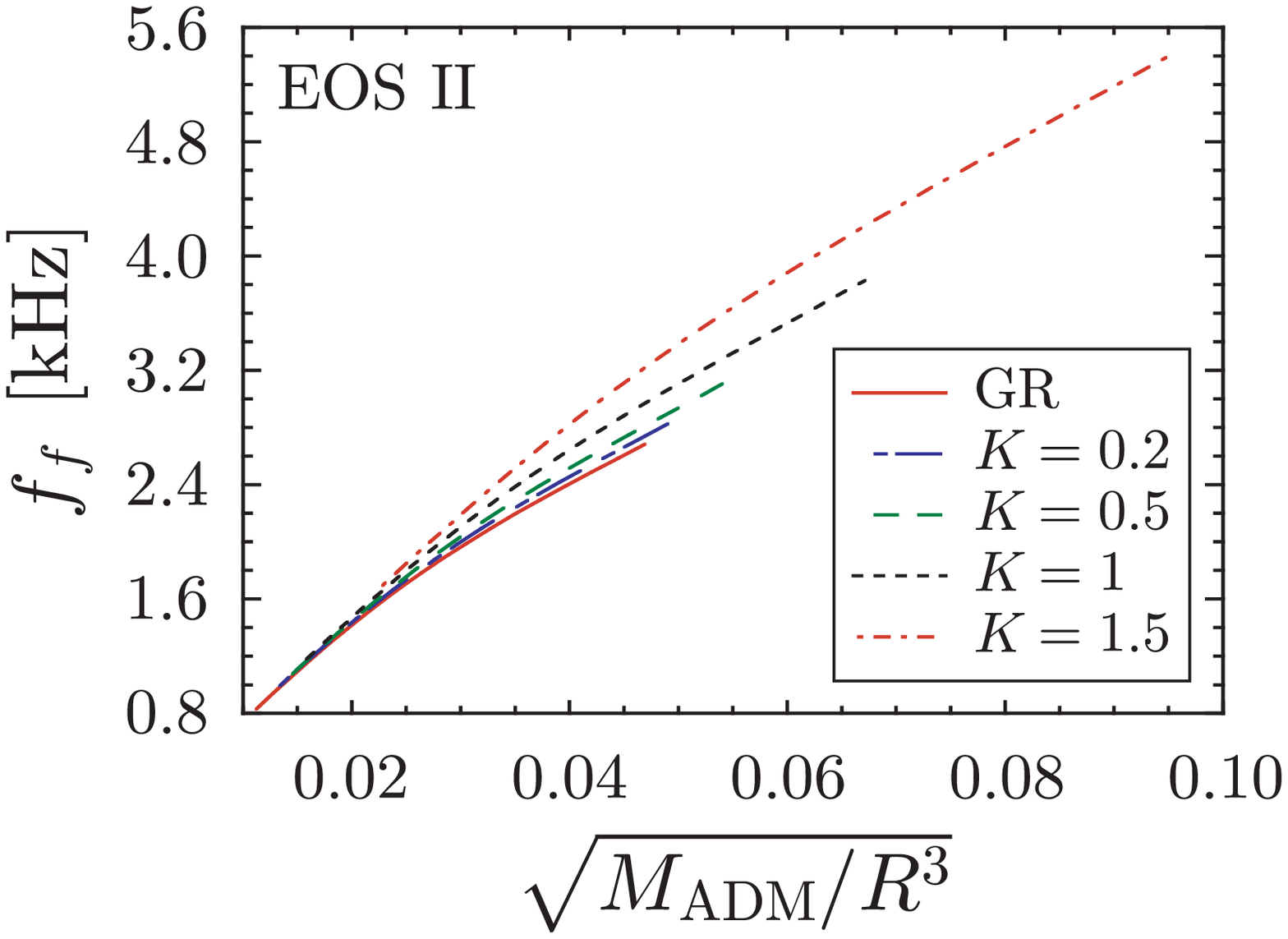} \\
\end{tabular}
\end{center}
\caption{%%
The frequency of $f$ mode as a function of the stellar average
density, $(M_{\rm ADM}/R^3)^{1/2}$, where $f_f$ is defined as $f_f\equiv \omega_f/(2\pi)$.
The solid line corresponds to the frequency in GR, while the other broken lines are
corresponding to the frequencies in TeVeS with several values of $K$.
Notice that the unit of average density is [1/km] in the geometrical unit, where $c=G=1$.
}%%
\label{fig:density-omega}
\end{figure}

With respect to the neutron star models shown in Sec. \ref{sec:II},
in this section we examine the stellar oscillations.
Especially, we focus on the stellar models whose central density is in the range
from $\tilde{\rho}_0=10^{14}$ g/cm$^3$ up to the value given the maximum ADM mass.
The stellar parameters with maximum ADM mass are summarized in Tables \ref{tab:EOSA}
and \ref{tab:EOSII}, where $\varphi_0$ and $z$ are central value of $\varphi$ and
surface redshift, respectively.
In general, the oscillation
spectrum is directly related to the stellar parameter, such as mass, radius and
EOS, but the frequencies of fundamental oscillation modes, i.e., $f$ modes, can be
connected to the stellar average density, $(M_{\rm ADM}/R^3)^{1/2}$. This reason
is physically explained by considering the relation between the sound speed and
the time that fluid perturbation needs to propagate across the star. Actually,
for the stellar models in GR, Andersson \& Kokkotas found the empirical formula
for the frequency of $f$ mode as a function of stellar average density \cite{Andersson1998}.
The $f$ mode frequencies for the stellar models in GR constructed with almost all EOS are subject to
this empirical formula. While, the frequencies of $f$ mode for the stellar models in TeVeS
with above two EOS are shown in Fig. \ref{fig:density-omega}.
The deviation from GR is clearly cognized for typical neutron stars and
depending on the value of parameter $K$, the frequencies become around 20 \% larger than those
expected in a general relativistic neutron star.
This can be an observable effect and one might distinguish the gravitational theory in strong
gravitational field by using the observations of gravitational waves.
%
%%%%%%%%%%%%%%%%%%%%%%%%%%%%%%%%%%%%%%%%%%%%%%%%%%%%%%%%%%%%%%%%%%%%%%%%%%%%%%%%%%%%%%%%%%%%%%%%
%  For EOS A,  3.28 kHz in GR while 4.02 kHz in TeVeS with K=1.5, which is 22.56 % deviation   %
%  For EOS II, 2.68 kHz in GR while 3.21 kHz in TeVeS with K=1.5, which is 19.78 % deviation   %
%%%%%%%%%%%%%%%%%%%%%%%%%%%%%%%%%%%%%%%%%%%%%%%%%%%%%%%%%%%%%%%%%%%%%%%%%%%%%%%%%%%%%%%%%%%%%%%%
%
%
%
%%%%%%%%%%%%%%%%%%%%%%%%%%%%%%%%%%%%%%%%%%%%%%%%%%%%%%%%%%%%%%%%%%%%%%
% Table 1
%%%%%%%%%%%%%%%%%%%%%%%%%%%%%%%%%%%%%%%%%%%%%%%%%%%%%%%%%%%%%%%%%%%%%%
\begin{table}[htbp]
\begin{center}
\leavevmode
\caption{Stellar parameters for models with EOS A and with maximum ADM mass,
where we choose that $k=0.03$ and $\varphi_c=0.003$.
}
\begin{tabular}{cc cc cc cc}
\hline\hline
 $K$ & $M_{\rm ADM}/M_\odot$ & $\tilde{\rho}_0$ [g/cm$^3$] & $R$ [km] & $\varphi_0$ & $M_{\rm ADM}/R$ & $z$  \\
\hline
% 0.1 & $1.61$ & $3.63 \times 10^{15}$ & $8.61$ & $6.49 \times 10^{-4}$ & $0.273$ & $0.486$  \\
 0.2 & $1.56$ & $3.54 \times 10^{15}$ & $8.36$ & $8.10 \times 10^{-4}$ & $0.271$ & $0.479$  \\
 0.5 & $1.42$ & $3.30 \times 10^{15}$ & $7.60$ & $1.26 \times 10^{-3}$ & $0.265$ & $0.459$  \\
 1.0 & $1.15$ & $3.00 \times 10^{15}$ & $6.16$ & $1.91 \times 10^{-3}$ & $0.255$ & $0.433$  \\
 1.5 & $0.81$ & $2.81 \times 10^{15}$ & $4.32$ & $2.48 \times 10^{-3}$ & $0.248$ & $0.414$  \\
 1.9 & $0.36$ & $2.70 \times 10^{15}$ & $1.92$ & $2.90 \times 10^{-3}$ & $0.242$ & $0.401$  \\
\hline\hline
\end{tabular}
\label{tab:EOSA}
\end{center}
\end{table}
%%%%%%%%%%
%
%
%%%%%%%%%%%%%%%%%%%%%%%%%%%%%%%%%%%%%%%%%%%%%%%%%%%%%%%%%%%%%%%%%%%%%%
% Table 2
%%%%%%%%%%%%%%%%%%%%%%%%%%%%%%%%%%%%%%%%%%%%%%%%%%%%%%%%%%%%%%%%%%%%%%
\begin{table}[htbp]
\begin{center}
\leavevmode
\caption{Stellar parameters for models with EOS II and with maximum ADM mass,
where we choose that $k=0.03$ and $\varphi_c=0.003$.
}
\begin{tabular}{cc cc cc cc}
\hline\hline
 $K$ & $M_{\rm ADM}/M_\odot$ & $\tilde{\rho}_0$ [g/cm$^3$] & $R$ [km] & $\varphi_0$ & $M_{\rm ADM}/R$ & $z$  \\
\hline
% 0.1 & $1.89$ & $2.45 \times 10^{15}$ & $10.62$ & $7.94 \times 10^{-4}$ & $0.261$ & $0.446$  \\
 0.2 & $1.84$ & $2.40 \times 10^{15}$ & $10.32$ & $9.38 \times 10^{-4}$ & $0.259$ & $0.441$  \\
 0.5 & $1.67$ & $2.25 \times 10^{15}$ &  $9.39$ & $1.35 \times 10^{-3}$ & $0.253$ & $0.425$  \\
 1.0 & $1.36$ & $2.06 \times 10^{15}$ &  $7.63$ & $1.96 \times 10^{-3}$ & $0.245$ & $0.403$  \\
 1.5 & $0.96$ & $1.92 \times 10^{15}$ &  $5.36$ & $2.51 \times 10^{-3}$ & $0.238$ & $0.385$  \\
 1.9 & $0.43$ & $1.80 \times 10^{15}$ &  $2.40$ & $2.91 \times 10^{-3}$ & $0.231$ & $0.370$  \\
\hline\hline
\end{tabular}
\label{tab:EOSII}
\end{center}
\end{table}
%%%%%%%%%%
%
%
%%%%%%%%%%%%%%%%%%%%%%%%%%%%%%%%%%%%%%%%%%%%%%%%%%%%%%%%%%%%%%%%%
%  FIGURE
%%%%%%%%%%%%%%%%%%%%%%%%%%%%%%%%%%%%%%%%%%%%%%%%%%%%%
% Figure 3
%
%%%%%%%%%%%%%%%%%%%%%%%%%%%%%%%%%%%%%%%%%%%%%%%%%%%%%
%%%%%%%%%%%%%%%%%%%%%%%%%%%%%%%%%%%%%%%%%

\begin{figure}[htbp]
\begin{center}
\begin{tabular}{cc}
\includegraphics[scale=0.45]{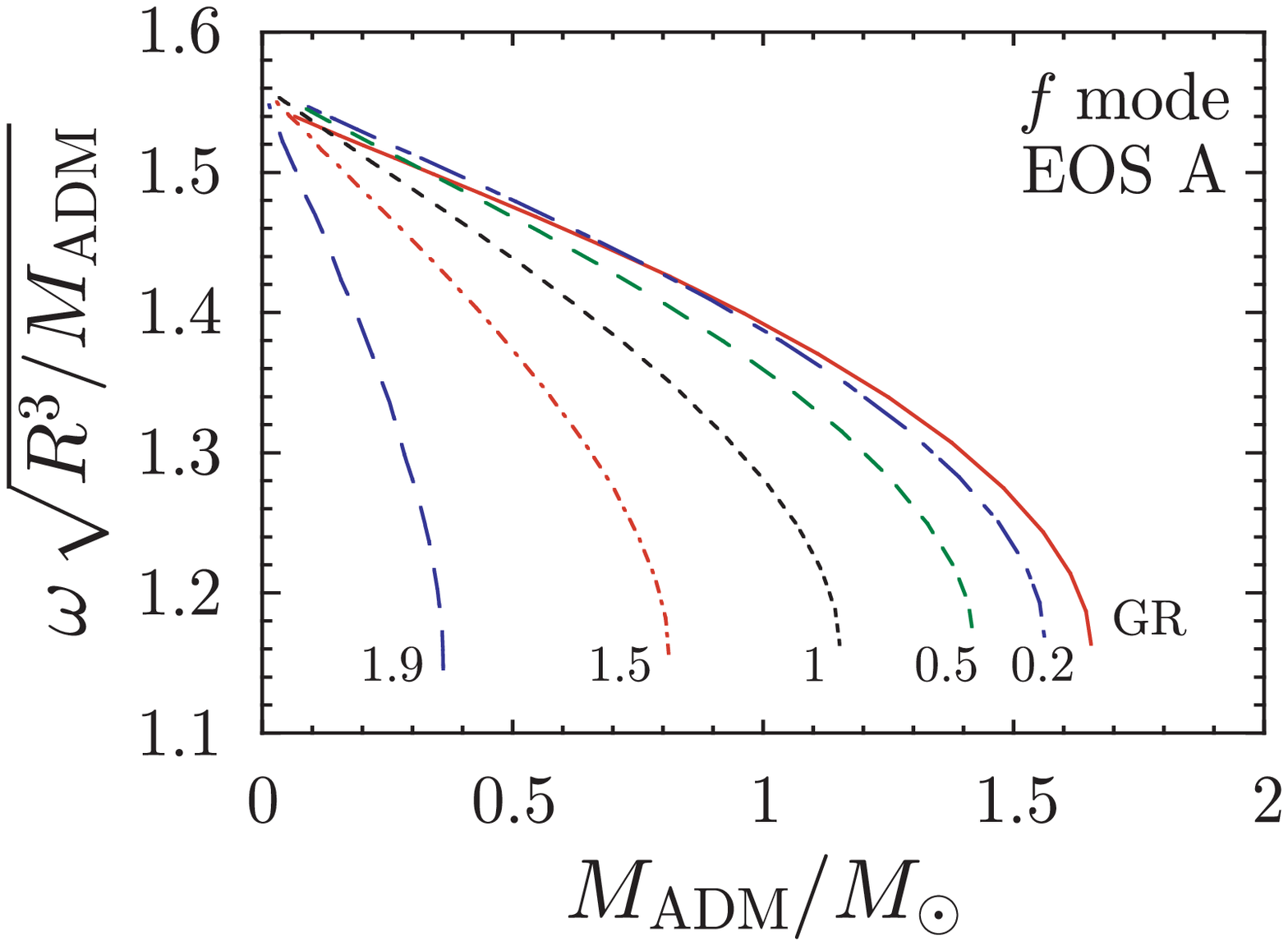} &
\includegraphics[scale=0.45]{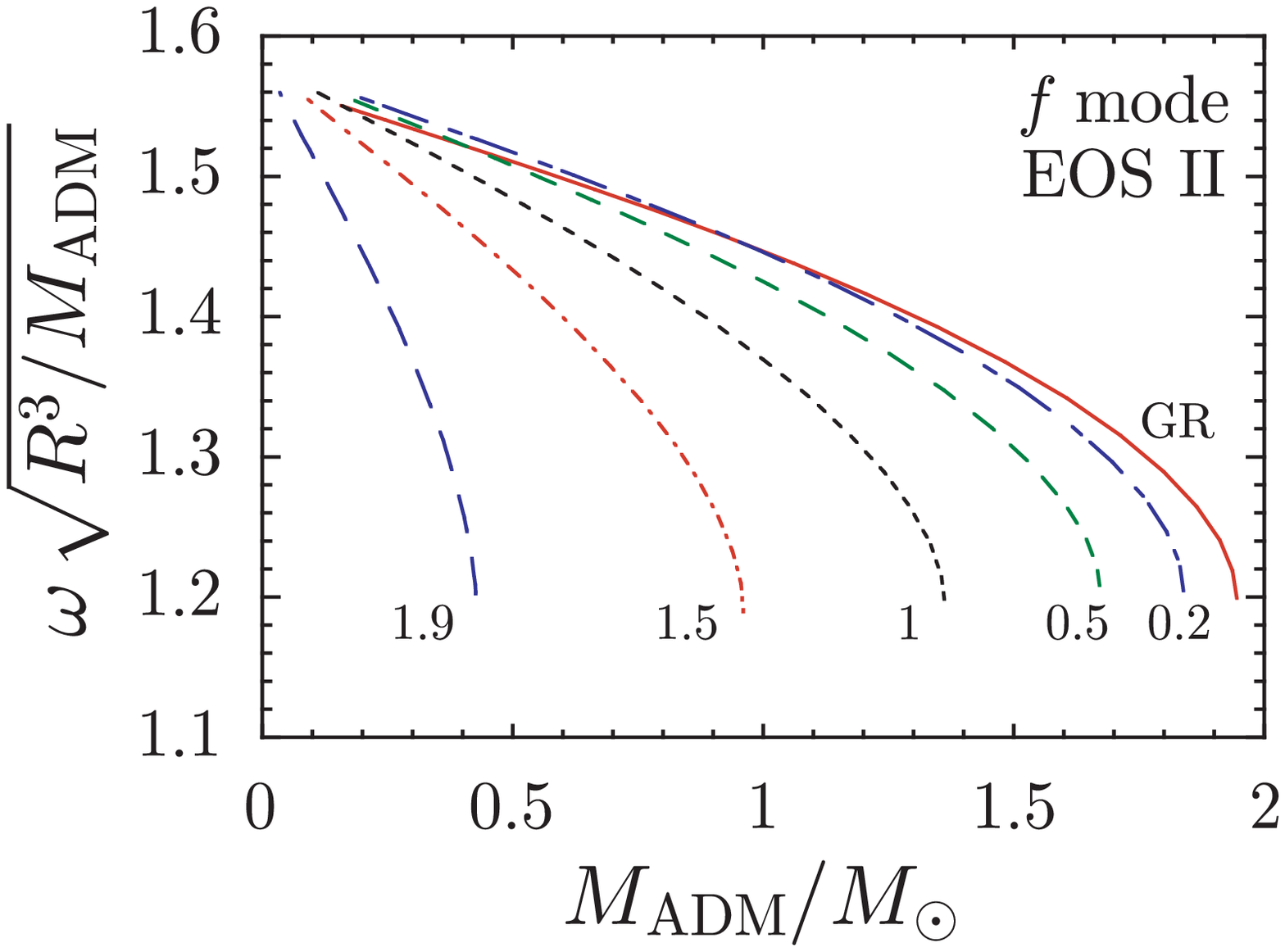} \\
\includegraphics[scale=0.45]{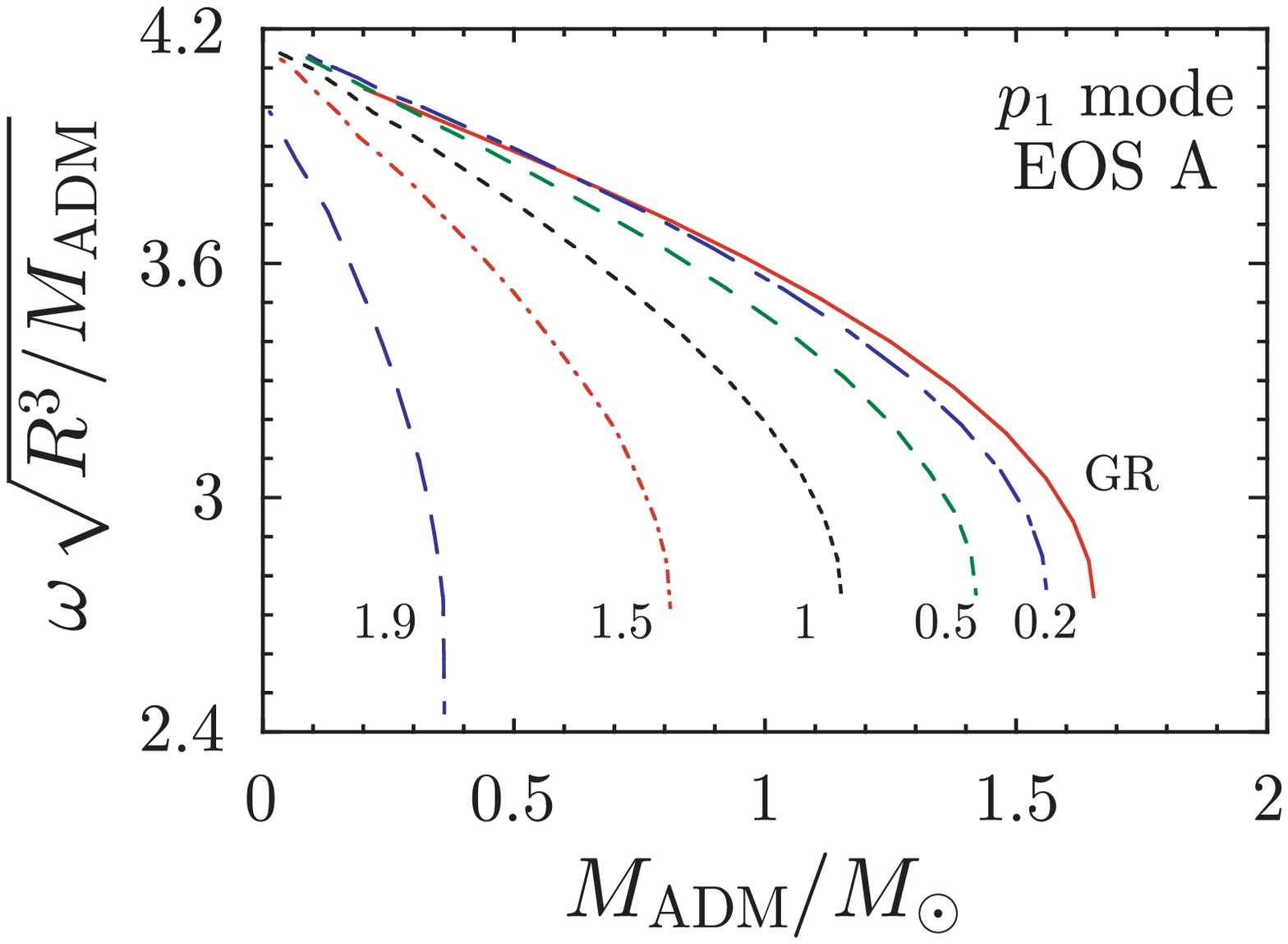} &
\includegraphics[scale=0.45]{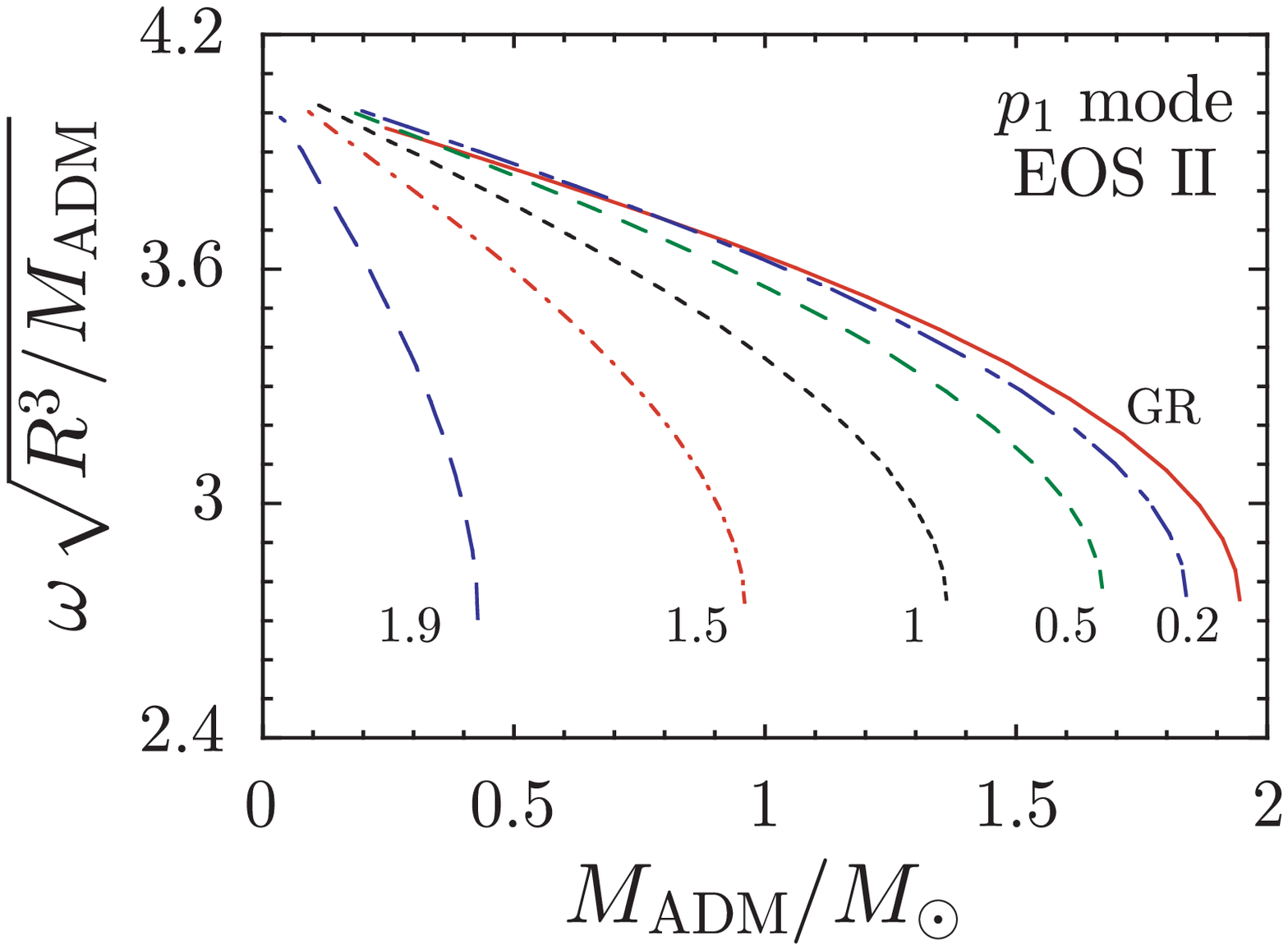} \\
\end{tabular}
\end{center}
\caption{%%
The normalized frequencies of first two fluid modes are plotted
as functions of the ADM mass, where upper and lower panels are
corresponding to $f$ and $p_1$ modes, respectively.
In these figures, the solid line corresponds to the frequency in GR,
while the other broken lines are
corresponding to the frequencies in TeVeS with several values of $K$.
}%%
\label{fig:M-omega}
\end{figure}

\begin{figure}[htbp]
\begin{center}
\begin{tabular}{cc}
\includegraphics[scale=0.45]{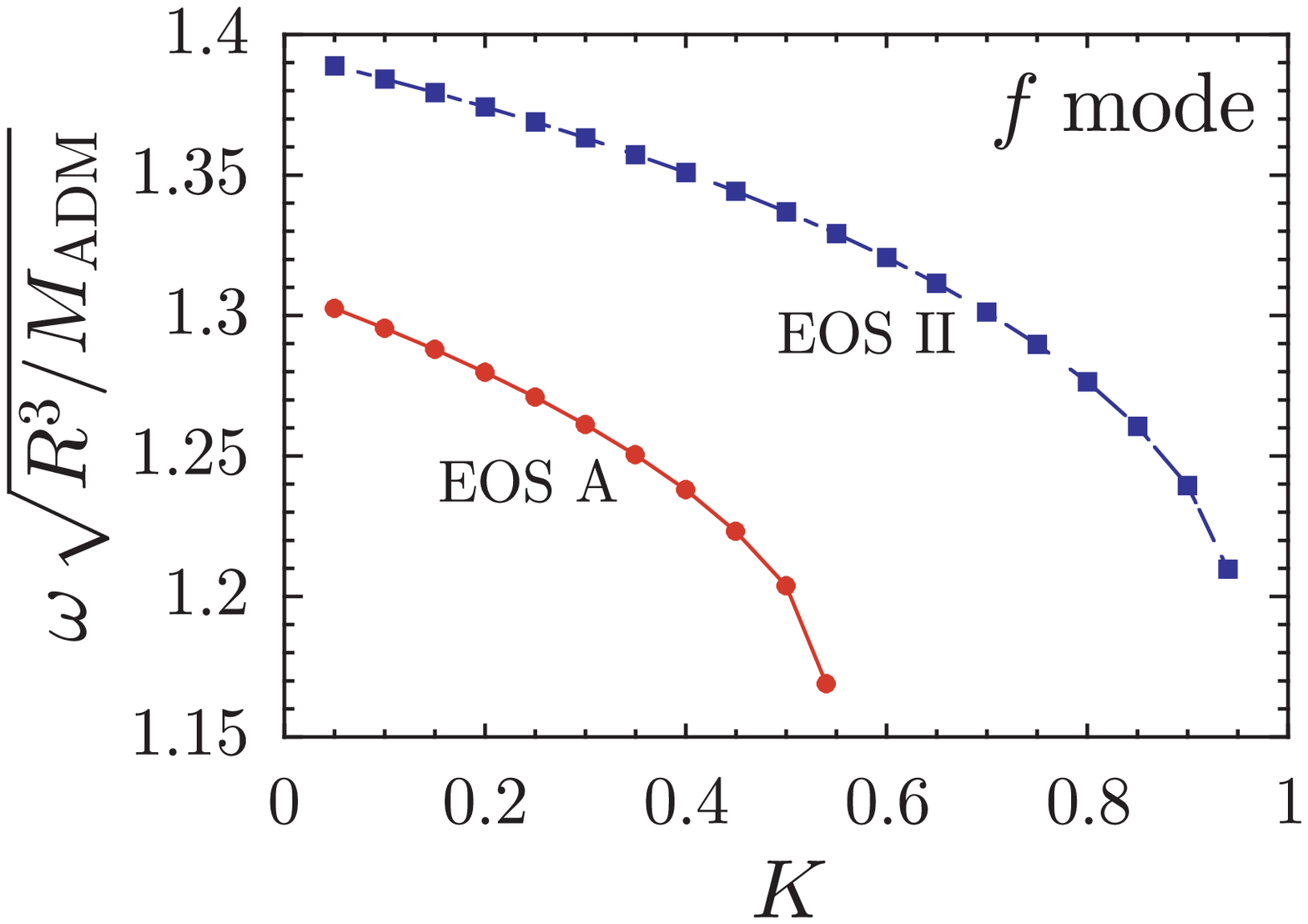} &
\includegraphics[scale=0.45]{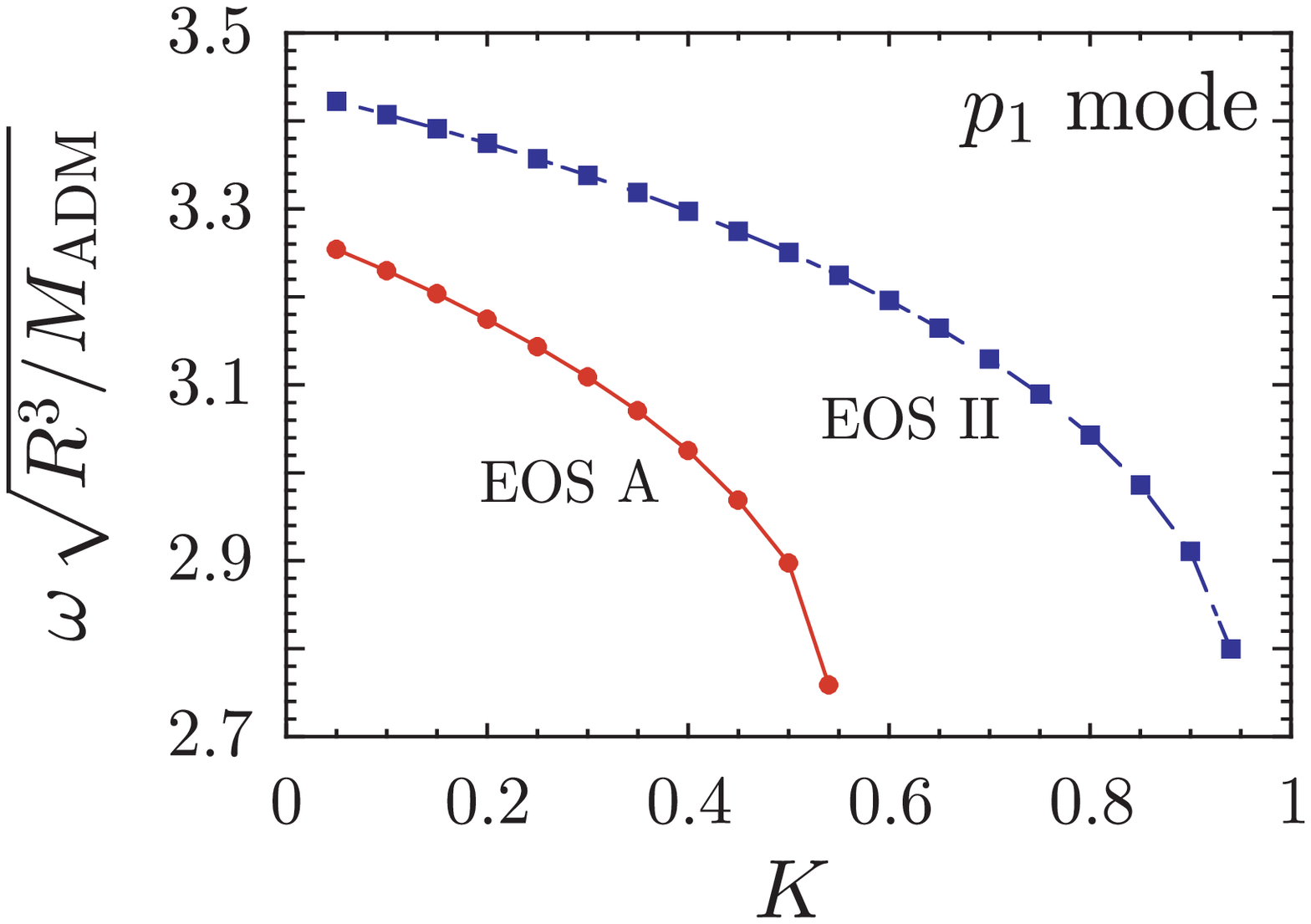} \\
\includegraphics[scale=0.45]{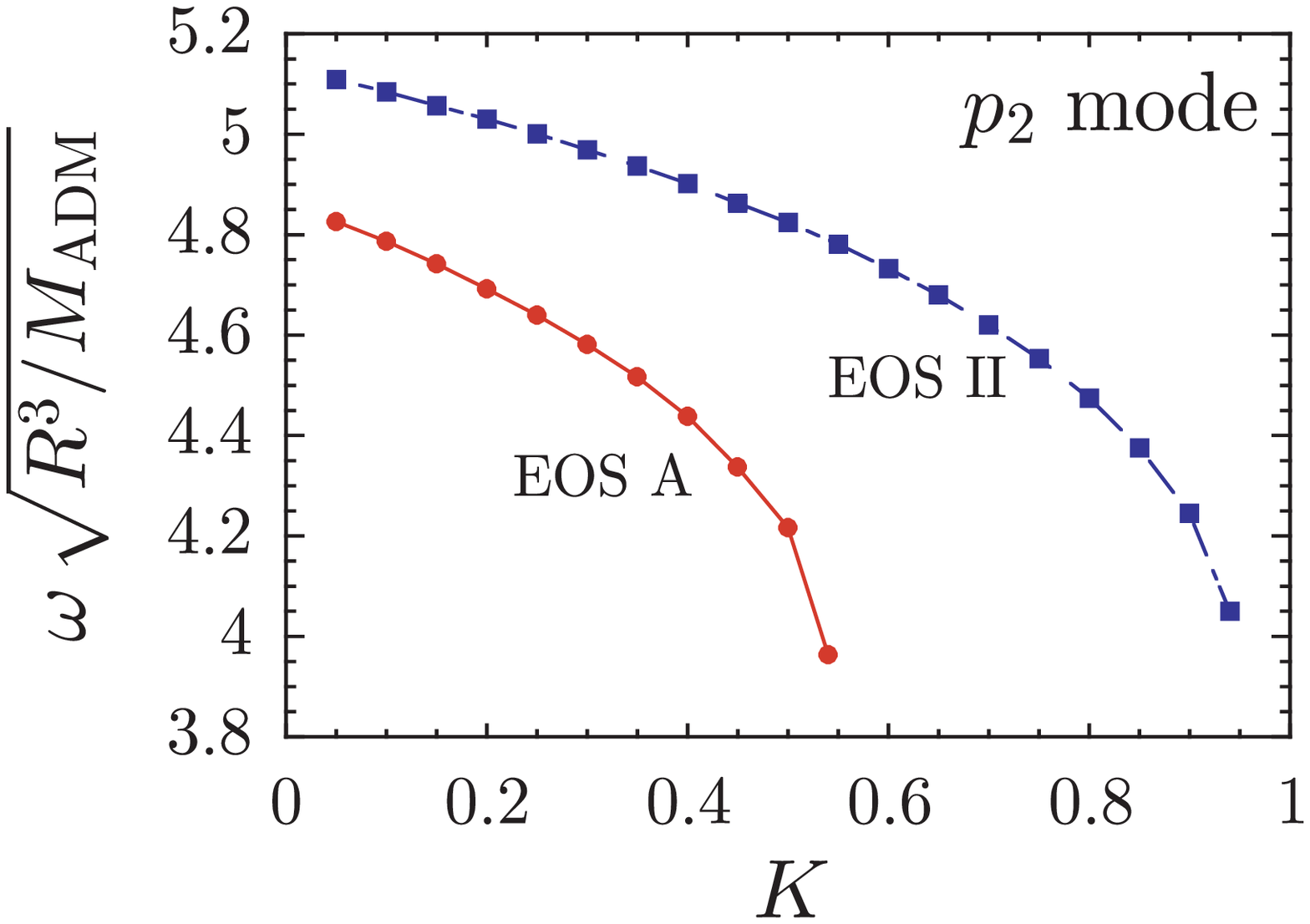} &
\includegraphics[scale=0.45]{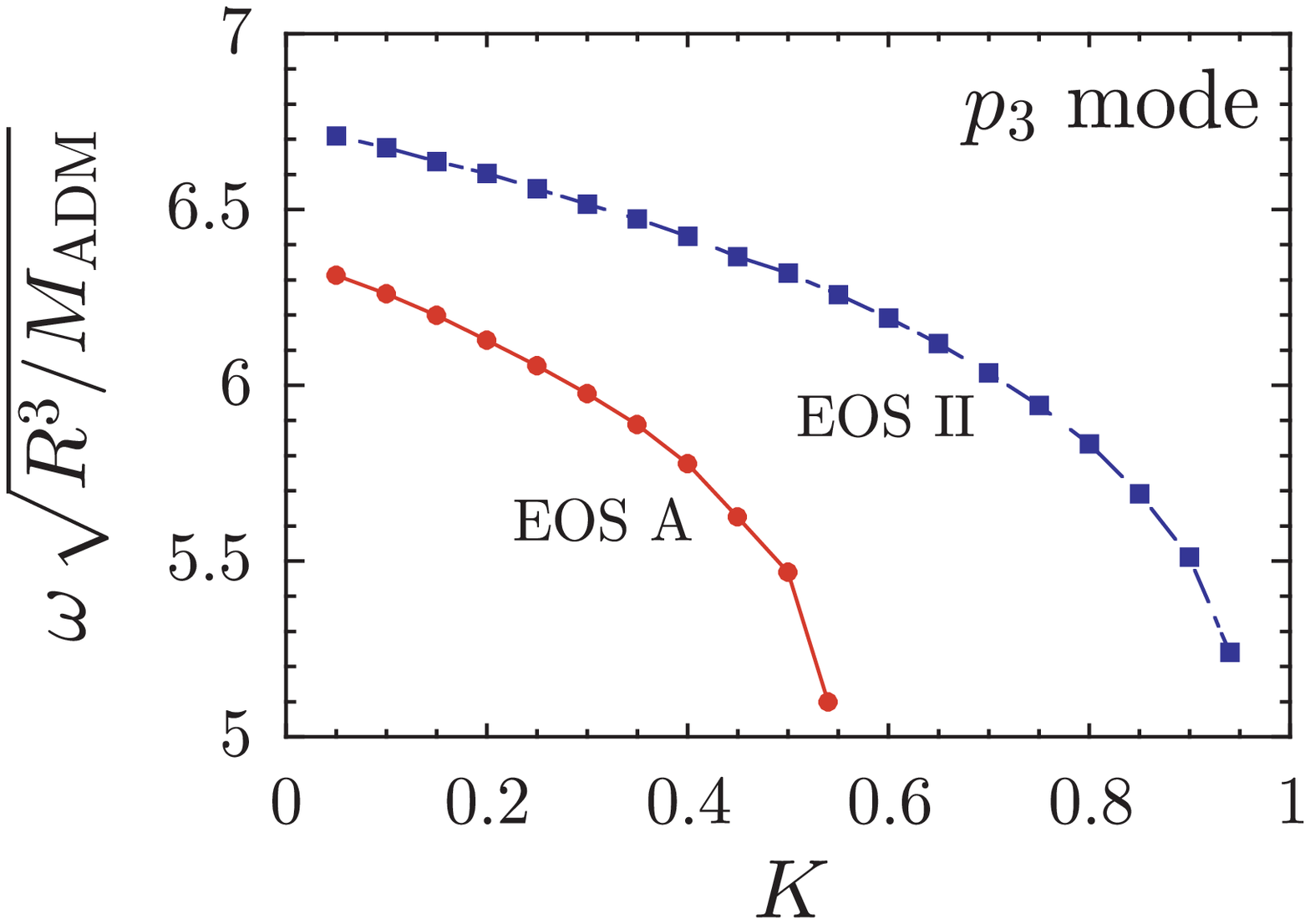} \\
\end{tabular}
\end{center}
\caption{%%
For the stellar models with $M_{\rm ADM}=1.4M_\odot$,
the normalized eigenvalues $\omega$ of the first few modes ($f$, $p_1$, $p_2$ and $p_3$) are shown
as functions of parameter $K$ with EOS A (solid lines) and EOS II (broken line). The values of
$k$ and $\varphi_c$ are fixed as $k=0.03$ and $\varphi_c=0.003$, respectively.
}%%
\label{fig:K-omega}
\end{figure}

The possibility to probe the gravitational theory by using observations of gravitational waves
can be also seen in Fig. \ref{fig:M-omega}, where we plot the normalized frequencies of $f$ and
$p_1$ modes as functions of ADM mass. In the figures, the solid line denotes the frequency in GR
while the other broken lines are corresponding to those in TeVeS with several values of $K$.
One can easily observe that the frequencies expected in TeVeS are quite different from those in GR.
Since this distinction results from the difference of gravitational theory, which creates
due to the presence of scalar field, observing more than one mode of gravitational wave
could tell us the existence of the scalar field. This statement might become more obvious by
seeing the dependence of frequencies of gravitational waves on the parameter $K$.
In Fig. \ref{fig:K-omega}, we plot the normalized frequencies of the first four modes, i.e.,
$f$, $p_1$, $p_2$ and $p_3$, as functions of parameter $K$, where the ADM masses are fixed
to be $1.4M_{\odot}$. The allowed maximum values of $K$ to produce the stellar models
with $M_{\rm ADM}=1.4M_\odot$ are $K=0.54$ for EOS A and $K=0.94$ for EOS II.
From these figures, we can see that the qualitative dependences of frequency
on the value of $K$ are independent from EOS and kinds of eigen-mode.
That is, the normalized frequencies of fluid modes are decreasing
as the value of $K$ becomes large. On the other hand, it is also found that the higher
overtone is quantitatively more sensitive against the value of $K$ than the lower modes.
This point can be seen in Table \ref{tab:ratio},
where we summarize the ratio of difference between the frequencies for
the stellar models with $K=0.05$ and with the allowed maximum values of $K$.
In other words, the frequencies of higher overtone is more helpful to distinguish TeVeS from GR
via the gravitational wave observations.
Anyway, through Figs. \ref{fig:M-omega} and \ref{fig:K-omega}, we can find that
with a help of observation of stellar mass, it is possible to probe
the gravitational theory in the strong-field regime by using observations of gravitational waves.

%%%%%%%%%%%%%%%%%%%%%%%%%%%%%%%%%%%%%%%%%%%%%%%%%%%%%%%%%%%%%%%%%
%  FIGURE
%%%%%%%%%%%%%%%%%%%%%%%%%%%%%%%%%%%%%%%%%%%%%%%%%%%%%
% Figure 4
%
%%%%%%%%%%%%%%%%%%%%%%%%%%%%%%%%%%%%%%%%%%%%%%%%%%%%%
%%%%%%%%%%%%%%%%%%%%%%%%%%%%%%%%%%%%%%%%%
%
%
%%%%%%%%%%%%%%%%%%%%%%%%%%%%%%%%%%%%%%%%%%%%%%%%%%%%%%%%%%%%%%%%%
%  FIGURE
%%%%%%%%%%%%%%%%%%%%%%%%%%%%%%%%%%%%%%%%%%%%%%%%%%%%%
% Figure 5
%
%%%%%%%%%%%%%%%%%%%%%%%%%%%%%%%%%%%%%%%%%%%%%%%%%%%%%
%%%%%%%%%%%%%%%%%%%%%%%%%%%%%%%%%%%%%%%%%
%
%
%%%%%%%%%%%%%%%%%%%%%%%%%%%%%%%%%%%%%%%%%%%%%%%%%%%%%%%%%%%%%%%%%%%%%%
% Table 3
%%%%%%%%%%%%%%%%%%%%%%%%%%%%%%%%%%%%%%%%%%%%%%%%%%%%%%%%%%%%%%%%%%%%%%
\begin{table}[htbp]
\begin{center}
\leavevmode
\caption{
%The ratio of difference between the frequencies for the stellar models with $K=0.05$
%and with the allowed maximum values of $K$, which are $K=0.54$ for EOS A and $K=0.94$ for EOS II.
The relative frequency change of each eigen-mode in Fig. \ref{fig:K-omega} defined as
$(\omega_{K=0.05}-\omega_{K=max})/\omega_{K=0.05}$, where $\omega_{K=0.05}$ and $\omega_{K=max}$
denote the frequencies for the stellar models with $K=0.05$
and with the allowed maximum values of $K$, respectively.
}
\begin{tabular}{cc cc c}
\hline\hline
 mode & & EOS A & & EOS II   \\
\hline
 $f$   & & $10.28\%$ & & $12.89\%$   \\
 $p_1$ & & $15.21\%$ & & $18.18\%$   \\
 $p_2$ & & $17.88\%$ & & $20.73\%$   \\
 $p_3$ & & $19.23\%$ & & $21.89\%$   \\
\hline\hline
\end{tabular}
\label{tab:ratio}
\end{center}
\end{table}
%%%%%%%%%%
%
%

%%%%%%%%%%%%%%%%%%%%%%%%%%%%%%%%%%%%%%%%%%%%%%%%%%%%%
%%%%%%%%%%%%%%%%%%%%%%%%%%%%%%%%%%%%%%%%%%%%%%%%%%%%%
\section{Conclusion}
\label{sec:V}
%%%%%%%%%%%%%%%%%%%%%%%%%%%%%%%%%%%%%%%%%%%%%%%%%%%%%
%%%%%%%%%%%%%%%%%%%%%%%%%%%%%%%%%%%%%%%%%%%%%%%%%%%%%

In this article, to examine the effect of the Tensor-Vector-Scalar (TeVeS) Theory on
the oscillation spectra of neutron stars, we have derived the perturbation
equations of neutron stars in TeVeS and calculated their eigen-frequencies.
Depending on the parameter of TeVeS, the frequencies of fundamental oscillation
could be off the well-known empirical formula in GR and they become lager
than those expected in GR. We can also see the deviation from GR in the frequencies
of higher overtones and they have stronger dependence on the parameter $K$
than the lower oscillation modes.
Since these imprints of TeVeS come from the presence of scalar field,
by using the observations of gravitational waves associated with the stellar oscillations,
it will be possible not only to distinguish the gravitational theory in the strong-field
regime, but also to probe the existence of the scalar field.

For simplicity, we assumed the Cowling approximation in this study,
which restricts our examination to only stellar oscillations. This means that we should
do more detailed study including the metric, vector and scalar fields perturbations.
%which will be seen near future somewhere.
Via these oscillations, we could obtain the additional information in the gravitational spectrum,
and combining those with results shown in this article would provide more accurate
constrains on the gravitational theory in the strong-field regime.
Furthermore the introduction of the stellar magnetic effect might be also important.
For example, recent observation of quasi-periodic oscillation in the giant flares
are believed to be related to the oscillations of strong magnetized neutron stars
\cite{Sotani2007,Sotani2008,Sotani2009}. Considering the magnetic effects, one might
be able to get the further constraint in the theory.

%\newpage
%%%%%%%%%%%%%%%%%%%%%%%%%%%%%%%%%%%%%%%%%%%%%%%%%%%%%%%%%%%%%%%%%%%%%%
\acknowledgments
%%%%%%%%%%%%%%%%%%%%%%%%%%%%%%%%%%%%%%%%%%%%%%%%%%%%%%%%%%%%%%%%%%%%%%

We thank K.D. Kokkotas for valuable comments.
This work was supported via the Transregio 7 ``Gravitational Wave Astronomy"
financed by the Deutsche Forschungsgemeinschaft DFG (German Research Foundation).

\appendix

%%%%%%%%%%%%%%%%%%%%%%%%%%%%%%%%%%%%%%%%%%%%%%%%%%%%%%%%%%%%%%%%%%%%%%%%%%%%%%%%%%%%%%%%%%%%%%%%%%%%%%%%%%%%%%%
%\section{}   % Appendix A
%\label{sec:appendix_1}
%%%%%%%%%%%%%%%%%%%%%%%%%%%%%%%%%%%%%%%%%%%%%%%%%%%%%%%%%%%%%%%%%%%%%%%%%%%%%%%%%%%%%%%%%%%%%%%%%%%%%%%%%%%%%%%

%%%%%%%%%%%%%%%%%%%%%%%%%%%%%%%%%%%%%%%%%%%%%%%%%%%%%%%%%%%%%%%%%%%%%%

\end{document}